\documentclass[12pt,prd,aps,amssymb,amsmath,tightenlines,showpacs]{article}
\usepackage[utf8]{inputenc}
\usepackage[a4paper,top=3cm,bottom=3cm,left=1.5cm,right=1.5cm]{geometry}
\usepackage{amssymb}
\usepackage{amsmath}
\usepackage{amsthm}
\usepackage{tipa}
\usepackage{latexsym} 
\usepackage{graphicx}
\usepackage{slashed}
\usepackage{bbold}
\usepackage{cancel}
\usepackage{comment}
\usepackage{color}
\usepackage{graphicx,epsfig,color}
\usepackage{comment}
\usepackage{cite}
\usepackage{tikz}
\usepackage[compat=1.1.0]{tikz-feynman}

\newcommand{\widesim}[2][1.5]{
	\mathrel{\underset{#2}{\scalebox{#1}[1]{$\sim$}}}
}
\newcommand{\be}{\begin{equation}}
\newcommand{\ee}{\end{equation}}
\newcommand{\bea}{\begin{eqnarray}}
\newcommand{\eea}{\end{eqnarray}}
\newcommand{\vv}{``}

\def\nn{\nonumber}

\begin{document}
	\graphicspath{{FIGURE/}}
	\topmargin=-1cm
	
\begin{center} 
	{\Large{\bf 
Naturalness and UV sensitivity in Kaluza-Klein theories
}}\\

	\vspace*{0.8 cm}
	
{Carlo Branchina$^{*,\,a,\,b}$ \let\thefootnote\relax\footnote{$^{*}$cbranchina@cau.ac.kr},
	Vincenzo Branchina$^{\dagger,\,c}$ \let\thefootnote\relax\footnote{$^\dagger$branchina@ct.infn.it} and Filippo Contino$^{\ddagger,\,c}$ \footnote{$^\ddagger$filippo.contino@ct.infn.it} }
\vspace*{0.4cm}

{${}^{a}$\it Department of Physics, Chung-Ang University, Seoul 06974, Korea}

\vspace*{0.4cm}

{${}^{b}$\it Laboratoire de Physique Th\'eorique et Hautes Energies (LPTHE), UMR 7589,\\
	Sorbonne Universit\'e et CNRS, 4 place Jussieu, 75252 Paris Cedex 05, France}

\vspace*{0.4cm}

{${}^c$\it Department of Physics, University of Catania, 
	and INFN,} \\
\vspace*{0.02cm}
{\it Via Santa Sofia 64, I-95123 
	Catania, Italy}\\

\vspace*{1 cm}

	{\LARGE Abstract}\\
	
\end{center}

\setcounter{footnote}{0} 

More than twenty years ago a paradigm emerged according to which a UV-insensitive Higgs mass $m_H$ and (more generally) a UV-insensitive Higgs effective potential $V_{1l}(\phi)$ are obtained from higher-dimensional theories with compact extra dimensions and Scherk-Schwarz supersymmetry breaking. Since then, these ideas have been applied to different models of phenomenological interest, including recent applications to the dark energy problem.  
A thorough analysis of the framework on which such a paradigm is based allows us to show that
a source of strong UV sensitivity for $m_H$ and $V_{1l}(\phi)$, intimately connected to the non-trivial topology of these models' spacetime, was missed. The usual picture of the Scherk-Schwarz mechanism and its physical consequences need to be seriously reconsidered.

\section{Introduction}
The Standard Model (SM) of particle physics has met with enormous successes, culminating ten years ago in the discovery of the Higgs boson \cite{ATLAS:2012yve,CMS:2012qbp}. However, several important issues (dark matter, matter-antimatter asymmetry, neutrino masses, flavour problem, strong CP-problem, ...) are left unsolved, and urge us towards a theory beyond the Standard Model (BSM). Among them the Naturalness and Hierarchy (NH) problems, and the way gauge theories should merge with gravity in a consistent unified theory, more generally the problem of a complete unification of forces.

Concerning the Naturalness and Hierarchy problems, traditional approaches such as supersymmetry and/or composite models have to cope with the experimental constraints coming from the LHC: the compositeness scale or the SUSY breaking scale, more generally, the scale at which new physics is expected to appear, should be in the TeV region, but no sign of new physics has been observed in this energy regime so far.
Other attempts, that suffer however from similar difficulties, are related to the possibility of lowering the fundamental scale of quantum gravity through large (compact) extra dimensions \cite{Arkani-Hamed:1998jmv,Antoniadis:1998ig} or warped dimensions, as in Randall-Sundrum models \cite{Randall:1999ee,Randall:1999vf}. Some other popular approaches are based on perturbative calculations, implemented with dimensional regularization and/or  perturbative RG equations, but a recent analysis has shown that the theoretical frameworks on which they are based are not sound (see \cite{Branchina:2022jqc,Branchina:2022gll} and references therein). Other lines of research have also been developed in the last years. These include: (i) the search for UV/IR mixing properties in a  quantum gravity/string framework (see for instance works on the role of modular invariance in the calculation of the Higgs mass in String Theory \cite{Abel:2021tyt}, and on the application of the Swampland arguments to the problem \cite{Cheung:2014vva,Ibanez:2017oqr,Craig:2019fdy}), that would require at least a partial breakdown of the EFT paradigm even below the Planck scale; (ii) cosmological selection mechanisms, originally inspired by the so-called relaxion of \cite{Graham:2015cka}, where different arguments are used with the aim of explaining the observed value of the weak scale in terms of preferential values determined by the cosmological evolution; (iii) the use of discrete symmetries, as opposed to those of the traditional approaches, started with the twin Higgs model proposed in \cite{Chacko:2005pe}; (iv) the possibility that the small value of the Higgs mass results from a self-organized approach to the critical point $m_H=0$\cite{Eroncel:2018dkg,Giudice:2021viw}.

As for the unification of forces, one of the most popular approaches
is inspired to the pioneering works of Kaluza \cite{Kaluza:1921tu} and Klein \cite{Klein:1926tv}, and
consists in considering theories with compact extra dimensions (sometimes related to string theory), most often with at least one supersymmetry. Let us take for instance a $5D$ theory with one compact circular dimension of radius $R$. 
Fourier expanding the $5D$ fields of the original action along the compact dimension $z$, the integration over $z$ results in an effective $4D$ action. The original $5D$ fields split into different representations of the $4D$ Lorentz group and give rise to infinite towers of $4D$ fields, the so-called Kaluza-Klein (KK) fields, one for each distinct representation they decompose into, with masses determined by the discrete values ${n}/{R}$ of the momentum along the compact dimension. 

More than twenty years ago the idea emerged  that some theories with compact extra dimensions could naturally provide finite (UV-insensitive) values for the masses of scalar particles\cite{Antoniadis:1998sd,Delgado:1998qr}, in particular of the Higgs boson\footnote{This was built on observations and calculations already put forward for other scopes in \cite{Antoniadis:1997ic,Antoniadis:1997zg}.}. This result was greeted with great enthusiasm, as it seemed to hint to a physically well motivated mechanism for the solution of the long-standing naturalness problem.   
In its original formulation, this mechanism was implemented by considering supersymmetric extensions of the SM, with one extra
dimension compactified on the orbifold $S_1/Z_2$, where $S_1$ is a circle of radius $R$ \cite{Antoniadis:1998sd,Delgado:1998qr}. The particle spectrum is given by KK-towers of states with masses depending on $n/R$ ($n = 0, 1, . . .$). Supersymmetry is broken through the Scherk-Schwarz mechanism\cite{Scherk:1978ta,Scherk:1979zr}, and the masses of the KK fields are identified as ($q_{b_a}$ and $q_{f_a}$ are the R-symmetry charges or the periodicities of the boson and fermion fields respectively):
\begin{align}
m^2_{b_a}=(n/R+q_{b_a})^2 \qquad \qquad
m^2_{f_a}=(n/R+q_{f_a})^2  \nn\,,
\end{align}
where \vv $a$" refers to each of the superpartner families, while $b$ and $f$ are for bosons and fermions respectively.

Indicating with $\phi$ one of the scalar fields, that could either be the $n=0$ mode of a scalar that lives in the $5D$ bulk, or a Higgs field living on a boundary brane, the resulting $4D$ one-loop effective potential $V_{1l}(\phi)$ is written as \cite{Antoniadis:1998sd,Delgado:1998qr}:
\begin{equation}\label{ftp}
	V_{1l}(\phi)=\frac{1}{2}\sum_a\sum_{i_a}(-1)^{\delta_{i_a,f_{_a}}} \sum_{n=-\infty}^{\infty} \int \frac{d^4p}{(2\pi)^4}\log\left(p^2+M_a^2(\phi)+\left(\frac nR+q_{_{i_a}}\right)^2\right),
\end{equation}
where $M^2_a(\phi)$ is the field dependent mass of each family that interacts with the field $\phi$, and $i_a\equiv b_{_a},f_{_a}$ indicates the boson or the fermion partner in the family.
Another typical identification of the $4D$ Higgs field from higher-dimensional theories with compact extra dimensions is obtained by considering gauge theories, where the extra components of the gauge fields serve as $4D$ scalars\cite{Manton:1979kb}. As it was observed in\cite{Hosotani:1983xw,Hosotani:1988bm}, and later re-derived and carefully analyzed in\cite{Antoniadis:2001cv}, in such a framework a finite one-loop Higgs potential (and thus a finite mass) is obtained with no need for supersymmetry. In the present work we limit ourselves to consider only supersymmetric models with Scherk-Schwarz breaking mechanism, and leave the analysis of the gauge-Higgs unification scenario, with the related Hosotani mechanism, to a forthcoming paper\cite{To Appear}.

Coming back to \eqref{ftp}, it is important to point out that  in the usual approach
the sum over $n$ and the integration over  $p$ are considered as independent operations. As we will see, this is a delicate point, and we will discuss this issue at length in due time. In\cite{Antoniadis:1998sd,Delgado:1998qr} it is stated that the sum over $n$ {\it must} be done {\it before} the integration over $p$ (as in finite temperature field theory\cite{Dolan:1973qd,Weinberg:1974hy,Kapusta:2006pm}), and the $p$ integration is performed introducing a hard momentum cutoff $\Lambda$. They find that, in addition to finite contributions, $V_{1l}(\phi)$ contains divergent terms as $\Lambda^5$, $M_a^2(\phi)\, \Lambda^3$ and $M_a^4(\phi)\,\Lambda$ (apart from $\Lambda^5$, these are all field dependent divergences). In supersymmetric theories those are canceled due to the presence of superpartners ($M^2_a(\phi)$ is the same both for $i_a=b_{_a}$ and $i_a=f_{_a}$), and a finite $V_{1l}(\phi)$ is obtained. 
We stress however that, once a hard cutoff $\Lambda$ for the integration over $p$ is introduced, the well-known finite $V_{1l}(\phi)$ of\cite{Antoniadis:1998sd,Delgado:1998qr} is obtained irrespectively of the order followed for the two operations (sum over $n$ and integration over $p$), as first hinted in\cite{Ghilencea:2001bv,Ghilencea:2001ug}. This point will be further investigated in the present work. Moreover we will see that there is a more subtle issue,  overlooked in the previous literature, ultimately related to the way the quantum fluctuations of the original $5D$ theory are treated. As the original theory lives in $5$ dimensions (one of which is compact), the dimensionally reduced $4D$ one-loop Higgs potential $V_{1l}(\phi)$ actually derives from the original $5D$ one-loop potential.  The consequences of this apparently obvious observation are profound, and will be carefully explored in this work.

A mechanism similar to the one proposed in\cite{Antoniadis:1998sd,Delgado:1998qr} was later implemented in slightly different supersymmetric models, with $M^2_a(\phi)=0$ and $q_{i_a}$ field-dependent, where the compactification was realized on $S_1/Z_2\times Z_2'$ \cite{Barbieri:2000vh} (see also\cite{Arkani-Hamed:2001jyj}). The calculation of the $4D$ potential $V_{1l}(\phi)$ yields the same expression as in \eqref{ftp}, with $M^2_a(\phi) \to 0$ and $q_{i_a}\to q_{i_a}(\phi)$. 
In\cite{Barbieri:2000vh,Arkani-Hamed:2001jyj} it is suggested that the finiteness of $V_{1l}(\phi)$ and $m_\phi^2$ is due to compactification, and this procedure is named ``KK-regularization".
It is also worth to stress that in the usual approach the need for considering the sum over the KK-modes all the way up to infinity is often presented as a requirement that comes from the higher-dimensional symmetries of the original theory ($5D$ Lorentz symmetry, $5D$ SUSY, \dots)\footnote{Sometimes the shift symmetry along the compact dimension is also evoked and presented as the reflection on the effective field theory of the modular invariance of the string theory in which the former should be embedded.}. 

The approach followed in\cite{Antoniadis:1998sd,Delgado:1998qr,Barbieri:2000vh,Arkani-Hamed:2001jyj} was challenged in\cite{Ghilencea:2001ug} (see also \cite{Kim:2001gk}), where it was hinted that, introducing a hard numerical cut ($L$) for the sum over the KK-modes, and a hard momentum cutoff ($\Lambda$) for the integration over $p$, divergences show up unless $L$ is taken much larger than $\Lambda$, more precisely $L/R\gg \Lambda$. They argued that under this condition the $\Lambda$-insensitive results of \cite{Antoniadis:1998sd,Delgado:1998qr,Barbieri:2000vh,Arkani-Hamed:2001jyj}, together with terms suppressed by powers of $\Lambda R/L$, are obtained. 
This work triggered a heated debate on the validity of the techniques used to obtain the finite $V_{1l}(\phi)$ and $m^2_H$. 
Some authors reacted to\cite{Ghilencea:2001ug} reproducing the finite results with the help of smooth regularizations\cite{Delgado:2001ex,Contino:2001gz}, suggesting that the hard cut $L$ used in\cite{Ghilencea:2001ug} for the sum  could be at the origin of the (unwanted) UV-sensitive terms. Accordingly,
it was argued that \vv the sharp truncation of the KK-modes spoils the tower structure of the 5D theory"\cite{Delgado:2001ex}.

More specifically, considering a model with interactions localized on an orbifold fixed point brane, for the calculation of $m_H^2$ in\cite{Delgado:2001ex} a \vv thick brane'' regularization was used. Indicating with $z$ the coordinate along the compact dimension, the  $\delta(z)$ function that localizes the interactions on the brane is smoothened with a gaussian $f(z;l_s)\equiv 1/\sqrt{2\pi l_s^2}\,e^{-z^2/2l_s^2}$, where $l_s$ is the thickness of the brane ($\lim_{l_s\to0}f(z;l_s)=\delta(z)$). Integrating over $z$, a smooth cutoff function $e^{-\pi^2\left(n/R+q\right)^2l_s^2}$ for the sum over each of the KK-towers is generated ($q$ generically indicates either $q_b$ or $q_f$). The authors find
the same finite results of\cite{Antoniadis:1998sd,Delgado:1998qr,Barbieri:2000vh,Arkani-Hamed:2001jyj}, and claim that their smooth procedure provides a solid derivation of them.
From a similar perspective, the issues raised in \cite{Ghilencea:2001ug} were also challenged in \cite{Contino:2001gz}. Considering again the one-loop correction to the Higgs boson mass, a Pauli-Villars regularization is implemented through the introduction of higher derivative terms.
This generates the smooth cutoff function\, {\small${\Lambda^4}\Big/{\Big[\Lambda^4+\left(p^2+\left(\frac{n}{R}+q\right)^2\right)^2\Big]}$}\, for each of the $q$ and each propagator, and (in their words) such a regularization \vv manifestly preserves supersymmetry". Once again the finite results of\cite{Antoniadis:1998sd,Delgado:1998qr,Barbieri:2000vh,Arkani-Hamed:2001jyj} are obtained.

It seemed that these and similar works\cite{Delgado:2001ex,Contino:2001gz,Barbieri:2001dm,Masiero:2001im} closed the debate, and that the community came to a general agreement in favour of the finiteness of $m_H^2$ and $V_{1l}(\phi)$, as for example later reported in the lecture notes\cite{Quiros:2003gg}, and more recently in\cite{Koren:2020pio}. Since then, these ideas have been implemented in several contexts and even in recent times they are actively applied when working on theories with compact extra dimensions\cite{Murayama:2012jh,Dimopoulos:2014aua,Craig:2014fka,GarciaGarcia:2015yfz,Antoniadis:2015chx,Abel:2016pyc,Delgado:2016vib,Abel:2016hgy,Abel:2017rch,Cohen:2018mgv,Matsui:2018tan,Buchmuller:2018eog,Fraser:2019ojt,Tsulaia:2022csz}. Very recently there has also been renewed interest in models with large extra dimensions in connection to a Swampland approach to the cosmological constant problem\cite{Montero:2022prj,Burgess:2023pnk}.

In this work we show that delicate and critical questions went (and are still going) unnoticed.  We will see that, when sufficient attention is paid to the correct implementation of the sum over $n$ and integration over $p$, the expectation of getting UV-insensitive results turns out not to be fulfilled. 
The goal of the present work is to investigate on these issues. We will see that the idea that compactification could provide a mechanism (complementary to supersymmetry) to obtain a finite Higgs boson mass needs to be deeply reconsidered. As a consequence, the sometimes evoked physical picture of the Scherk-Schwarz breaking as an \vv effective lower-dimensional" SUSY breaking, more precisely as an apparent non-local and thus finite breaking for low-energy observers that cannot resolve the additional dimension, needs to be revisited too. In this respect, it is worth to stress that in their original work Scherk and Schwarz only use the extra dimensions as a trick/tool to generate mass splittings in the context of \vv dimensional reduction", but do not give them any physical reality. The question that needs to be reconsidered then concerns the nature of what is usually named Scherk-Schwarz SUSY breaking mechanism when the compact extra dimensions are taken as physical.

The rest of the paper is organized as follows. In section 2 we pave the way to our analysis. Starting from a $5D$ theory with one compact dimension, we review the usual way to derive $V_{1l}(\phi)$, 
and recover the usual results\cite{Antoniadis:1998sd,Delgado:1998qr,Barbieri:2000vh,Arkani-Hamed:2001jyj,Delgado:2001ex,Contino:2001gz,Barbieri:2001dm}. Starting with the $5D$ action of the original theory, in section 3 we specify and set up the physical and mathematical ingredients needed for a careful derivation of the (dimensionally reduced) $4D$ Higgs potential $V_{1l}(\phi)$. In section 4 we calculate $V_{1l}(\phi)$ with the help of the  Euler-McLaurin formula, that allows us to obtain the full result for $V_{1l}(\phi)$ for the first time. We use a hard cutoff that we implement in two different ways: in section 4.1 we introduce a numerical cut $L$ for the sum over $n$, and a spherical cutoff $\Lambda$ for the remaining four components of the loop momentum; in section 4.2 a spherical hard cutoff for the whole $5D$ loop momentum is considered.
Section 5 is devoted to the study of the problem through the introduction of a smooth cutoff. In section 5.1 we calculate $V_{1l}(\phi)$ performing the infinite sum over the KK-modes (with no cut as in section 2), and introduce a smooth cutoff function for the integration over $p$, getting the well-known finite result. In section 5.2 we introduce a smooth cutoff function for the whole $5D$ momentum $(p,n/R)$, getting this time the results of section 4.2. In section 6 we compare our results with those of previous literature. Section 7 is for the conclusions.

\section{One-loop potential. Infinite KK-tower sum}
\label{KK regularization calculation}
To introduce the tools for our work and pave the way to our analysis, in the present section we review the usual way a $4D$ one-loop potential $V_{1l}(\phi)$ is derived from a higher-dimensional theory with compact extra dimensions. 

As specified in the Introduction, we limit ourselves to the case of higher-dimensional supersymmetric theories with compact dimensions and SUSY broken by the Scherk-Schwarz mechanism\cite{Scherk:1978ta,Scherk:1979zr}. The case of higher-dimensional gauge theories with compact extra dimensions is left for a future publication\cite{To Appear}. For our illustrative purposes it is sufficient to focus on $5D$ theories
(the results are easily generalized to $4+n$ dimensions).
Let us consider for example the following (euclidean) action in $4+1$ dimensions ($a= 1,\dots,5$), 
\begin{equation}
\label{complex scalar action}
\mathcal{S}_{_{(5)}}[\widehat\Phi,\widehat\chi]=\int\mathrm{d}^{5}y\left\{\frac{1}{2}\partial_a \widehat\Phi\,\partial^a\Phi+\partial_a \widehat \chi \,\partial^a \widehat\chi^\dagger+\widehat M^2(\widehat\Phi)\widehat\chi\widehat\chi^\dagger \right\},
\end{equation}
where $\widehat\Phi(y)$ and $\widehat\chi(y)$ are $5$-dimensional scalar fields (real and complex  respectively), with $y\equiv(x,z)$, $z$ being the compact spatial coordinate along $S^1$ and $x$ the spacetime coordinate on $\mathbb R_4$ ($\mathcal M_4$ before euclideanization). Moreover, $\widehat M^2(\widehat\Phi)\equiv\widehat m^2+f(\widehat\Phi)$, where $\widehat m$ is the $\widehat \chi$ mass and $f(\widehat \Phi)\,\widehat\chi\widehat\chi^\dagger$ the interaction between $\widehat \Phi$ and $\widehat \chi$. As for the periodicities of the fields along the compact dimension, for $\widehat \Phi$ we have
\begin{align} 
\widehat \Phi(x,z+2\pi R)= \widehat \Phi(x,z)\,, \end{align}
while for the complex scalar field $\widehat \chi$ we consider the non-monodromy  
\begin{align} \label{nontrivialbc}
\widehat \chi(x,z+2\pi R)=e^{i2\pi Rq_\chi}\widehat \chi(x,z)
\end{align}
allowed by the existence of an internal symmetry, whose charge is $q_\chi$.
Expanding the fields in Fourier components:
\begin{align}\label{hosotani}
\widehat \Phi (x,z)&=\frac{1}{2\pi R}\sum_{n=-\infty}^{+\infty}\int\frac{d^4p}{(2\pi)^4}\,\widehat \Phi_{n,p}\,e^{i\left(p\cdot x+\frac{n}{R}z\right)}\equiv\frac{1}{\sqrt{2\pi R}}\sum_{n=-\infty}^{+\infty} \phi_n(x)e^{i\frac{n}{R}z}  \nn \\
\widehat \chi (x,z)&=\frac{1}{2\pi R}\sum_{n=-\infty}^{+\infty}\int\frac{d^4p}{(2\pi)^4}\,\widehat \chi_{n,p}\,e^{i\left(p\cdot x+\left(\frac{n}{R}+q_{\chi}\right)z\right)}\equiv\frac{1}{\sqrt{2\pi R}}\sum_{n=-\infty}^{+\infty} \chi_n(x)e^{i(\frac{n}{R}+q_\chi)z}\,,
\end{align}
where the $4D$ fields $\phi_n(x)$ are defined by (a similar equation holds for $\chi_n(x)$)
\begin{equation}
\phi_n(x) \equiv \frac{1}{\sqrt{2\pi R}}\int\frac{d^4p}{(2\pi)^4}\,\widehat \Phi_{n,p}\,e^{ip\cdot x}\,.
\end{equation}
In realistic phenomenological applications, additional fermion and boson fields (with supersymmetry explicitly implemented), and possibly different compactifications (typically orbifolds), are considered.
The Higgs field is usually identified either  with the $n=0$ mode $\phi_0(x)$ of the KK-tower of a scalar field $\widehat \Phi(y)$\cite{Delgado:1998qr,Antoniadis:1998sd,Barbieri:2000vh,Arkani-Hamed:2001jyj}, or with a $4D$ scalar field $\varphi(x)$ confined on a brane placed at the orbifold fixed point\cite{Arkani-Hamed:2001jyj,Delgado:2001ex}.

In the next section we will focus our attention on the question of considering an {\it infinite sum} over $n$ and the whole $\mathbb R_4$ space for the integration over $p$. For the time being, we start the usual calculation of $V_{1l}(\phi)$ inserting the expansions \eqref{hosotani} in \eqref{complex scalar action}, and
performing the integration over the compact dimension $z$, thus ending up with a dimensionally reduced $4D$ action for the KK-fields $\phi_n(x)$ and $\chi_n(x)$. The $4D$ one-loop Higgs potential $V_{1l}(\phi)$ is then calculated summing up the loop contributions\cite{Coleman:1973jx} from the infinitely many fields that appear in this action.
When the Higgs field is identified with the zero KK-mode of a $5D$ scalar field, as $\widehat \Phi$ in \eqref{complex scalar action} (see for instance\cite{Antoniadis:1998sd,Delgado:1998qr}), or with a brane field localized on an orbifold fixed point (see for instance\cite{Arkani-Hamed:2001jyj,Delgado:2001ex}), $V_{1l}(\phi)$ takes the form (the index $i$ runs over the families of bosons ($b$) and fermions ($f$) of the considered model)
\begin{equation}\label{EP1}
V_{1l}(\phi)=\frac{1}{2}\sum_i(-1)^{\delta_{if}} \sum_{n=-\infty}^{\infty} \int \frac{d^4p}{(2\pi)^4}\log\left(p^2+M^2(\phi)+\left(\frac nR+q_i\right)^2\right)\,,
\end{equation}
while in other supersymmetric theories, with $S^1/Z_2\times Z_2'$ compactification and mass mixing\cite{Barbieri:2000vh}, it is given by
\begin{equation}\label{EP2}
V_{1l}(\phi)=\frac{1}{2}\sum_i(-1)^{\delta_{if}} \sum_{n=-\infty}^{\infty} \int \frac{d^4p}{(2\pi)^4}\log\left(p^2+\left(\frac nR+q_i(\phi)\right)^2\right)\,.
\end{equation}

For our illustrative purposes, we do not need to concentrate on details of the (orbifold) compactifications considered, or on specific phenomenologically viable models that lead to \eqref{EP1} or \eqref{EP2}. Dwelling too much into them would obscure the message/focus of this work: the UV behaviour of $V_{1l}(\phi)$ can be studied from \eqref{EP1} and \eqref{EP2}, independently of the model from which either of them is derived.

In\cite{Barbieri:2000vh,Arkani-Hamed:2001jyj} supersymmetric theories that lead to a potential of the kind \eqref{EP2} (or directly to the corresponding one-loop contribution to the Higgs mass) were considered. The authors, performing {\it first} the sum over the infinite tower of KK-modes, and {\it successively} the infinite integration over the four-momentum $p$, obtain for \eqref{EP2} a finite result, the same being obviously true for $m_H^2$. They dubbed this procedure \vv KK-regularization". It was later observed (and we will see it below) that for models described by \eqref{EP2} supersymmetry is not necessary, as the boson and fermion contributions are separately finite\cite{Ghilencea:2001ug}. 
When the same calculation is done for models described by \eqref{EP1}, $V_{1l}(\phi)$ turns out to be finite only if the model is supersymmetric: finite contributions are obtained when the superpartners are combined together, but each single tower gives rise to divergences proportional to powers of $M^2(\phi)$. 

Let us move now to the explicit calculation of $V_{1l}(\phi)$.
We begin with \eqref{EP1} and consider first a generic bosonic contribution $V_{1l}^b(\phi)$, subtracting a normalization term $\log\, (p^2+ \frac{n^2}{R^2})$  to $\log\,(p^2+M^2+(\frac nR+q_{_b})^2)$.
Following the strategy usually adopted in the literature (see the comments above Eq.\,\eqref{EP1}), we perform separately the sum over $n$ and the integral over $p$, invert in \eqref{EP1} the sum with the integral, and carry out the former first. A simple look to \eqref{EP1} shows that, independently of getting or not a finite result for $V_{1l}(\phi)$, this calculation is possible only if a cutoff $\Lambda$ in the momentum integral is introduced (whatever order we consider in performing the sum and the integral).
With the help of the Schwinger identity (the upper case $\Lambda$ indicates that the integral over $p$ is perfomed with a cut-off $\Lambda$), we can then write:
\begin{equation}\label{ellipticthetaintp1}
	V^{b}_{1l}(\phi)=-\frac{1}{2} \int^{\Lambda} \frac{d^4p}{(2\pi)^4}\int_0^\infty \frac{ds}{s}\sum_{n=-\infty}^{\infty} \left[e^{-s(R^2(p^2+M^2)+(n+Rq_{_b})^2)}-e^{-s\,(p^2R^2+n^2)}\right].
\end{equation}
As the above series is uniformly convergent, we also exchanged the integral over $s$ with the sum over $n$. With the help of the Poisson resummation formula we obtain\footnote{Note for the expert reader. As we introduced the cutoff $\Lambda$ for the momentum integration, the lower extreme of the proper time integral need not to be replaced with an UV cutoff, but should be kept zero, since the Schwinger parametrization is here used only as an identity.}
\begin{equation}
\label{V after Poisson}
V^{b}_{1l}(\phi)=-\frac{\sqrt \pi}{2} \int^{\Lambda} \frac{d^4p}{(2\pi)^4}\int_0^\infty \frac{ds}{s^{3/2}} \left[ \vartheta _3\left(\pi  R q_{_b},e^{-\frac{\pi ^2}{s}}\right) e^{-s R^2 \left(p^2+M^2\right)}-  \vartheta _3\left(0,e^{-\frac{\pi ^2}{s}}\right)e^{-s R^2p^2}\right],
\end{equation}
where 
\begin{align}
\label{theta3}
\vartheta_3(x,y)=1+2 \sum_{k=1}^{\infty} \cos(2k x) y^{k^2}. 
\end{align}
Taking first for both $\vartheta_3$ in \eqref{V after Poisson} only the first term in the right hand side of \eqref{theta3}, we get
\begin{align}\label{ints}
-\frac 12 \int_0^\infty \frac{ds}{s^{3/2}} \sqrt{\pi } \left[ e^{-s R^2 \left(M^2+p^2\right)}- e^{-s R^2p^2}\right]= \pi R \left(-p+\sqrt{M^2+p^2}\right)\,,
\end{align}
and integrating\footnote{It is worth to note that \eqref{divm} is the result one would obtain for the one-loop potential in $5$ non-compact dimensions ($\mathcal M^5$, or $\mathbb R^5$ after euclideanization, where, being the spacetime simply connected, there is obviously no room for any $q$), when the integration over $p_5$ is performed in the whole range $]-\infty,\infty[$ (i.e.\,with loop integral $\int_{-\infty}^\infty dp_5\int^\Lambda d^4p$). We also note that it differs from the result that would be obtained in $\mathcal M^5\,(\mathbb R^5)$ with a more uniform cutoff, say $p^2+p_5^2\le\Lambda^2$, only for the coefficients in front of the divergent terms.} finally over $p$ we have
\begin{align}\label{divm}
& \int^{\Lambda}\frac{d^4p}{(2\pi)^4} \pi  R \left(-p+\sqrt{M^2+p^2}\right)\nonumber\\
&=-\frac{\Lambda ^5R}{40 \pi }+\frac{M^5 R}{60 \pi }+\frac{\Lambda ^2 M^2 R \sqrt{\Lambda ^2+M^2}}{120 \pi }+\frac{\Lambda ^4 R \sqrt{\Lambda ^2+M^2}}{40 \pi }-\frac{M^4 R \sqrt{\Lambda ^2+M^2}}{60 \pi }\nonumber\\
& \sim R \left(\frac{\Lambda ^3 M^2}{48 \pi }-\frac{\Lambda M^4}{64 \pi }+\frac{M^5}{60 \pi }+O\left(\frac{1}{\Lambda}\right)\right),
\end{align}
where in the last line we expanded the result for $M/\Lambda\ll 1$. 

Considering now the second term in the right hand side of \eqref{theta3}, and inserting it in \eqref{V after Poisson}, we note that both series in $k$ are uniformly convergent, so that we can treat each of the two integrals over $s$ separately. Again we can exchange the sum over $k$ with the integration over $s$, and picking up from \eqref{V after Poisson} only the contribution coming from the first $\vartheta_3$ we obtain
\begin{align}\label{risintdsp}
\int_0^{\infty}\frac{ds}{s^{3/2}}\sqrt{\pi } e^{-s R^2 \left(M^2+p^2\right)}\sum_{k=1}^\infty e^{-\frac{\pi ^2 k^2}{s}} \cos (2 \pi  k Rq_{_b})=\sum_{k=1}^\infty\frac{\cos (2 \pi  k Rq_{_b}) e^{-2 \pi  k R \sqrt{M^2+p^2}}}{k}.
\end{align}
Inserting \eqref{risintdsp} in  \eqref{V after Poisson}, and exchanging the sum over $k$ with the integration over $p$ we have 
\begin{align}\label{litium}
&-\sum_{k=1}^{\infty}\int^{\Lambda}\frac{d^4p}{(2\pi)^4}\frac{\cos (2 \pi  k Rq_{_b}) e^{-2 \pi  k R \sqrt{M^2+p^2}}}{k} \nonumber\\
&=\sum_{k=1}^{\infty}\left[-\frac{e^{-2 \pi k M R} (2 \pi k M R (2 \pi k M R+3)+3) \cos (2 \pi  k Rq_{_b})}{64 \pi ^6 k^5 R^4}+\frac{e^{-2 \pi  k R \Lambda}\Lambda ^3 \cos (2 \pi  k Rq_{_b})}{2 \pi  k^2 R}+\dots\right],
\end{align}
where the second term in the square brackets, as well as the terms indicated with the dots, are suppressed in the $\Lambda \to \infty$ limit. 
Finally, from the second $\vartheta_3$ function in \eqref{V after Poisson}, we obtain
\begin{align}\label{finterm}
\sum_{k=1}^\infty\int^{\Lambda} \frac{d^4p}{(2\pi)^4}\int_0^{\infty}\frac{ds}{s^{3/2}}\sqrt{\pi } e^{-s R^2 p^2} e^{-\frac{\pi ^2 k^2}{s}} =\sum_{k=1}^\infty\int^{\Lambda} \frac{d^4p}{(2\pi)^4}\frac{e^{-2 \pi  k R p}}{k} \simeq\sum_{k=1}^\infty \frac{3}{64 \pi ^6 k^5 R^4}=\frac{3 \zeta (5)}{64 \pi ^6 R^4}\,,
\end{align}
where $\Lambda$ suppressed terms are ignored. 
Adding together \eqref{divm}, \eqref{litium} and \eqref{finterm} (and neglecting all the $\Lambda$ suppressed terms), for $V^{b}_{1l}(\phi)$ we get\cite{Antoniadis:1998sd,Delgado:1998qr}
\begin{align}\label{bosonfinalp}
V^{b}_{1l}(\phi)&=R \left(\frac{\Lambda ^3 M^2}{48 \pi }-\frac{\Lambda M^4}{64 \pi }+\frac{M^5}{60 \pi }\right) +\frac{3 \zeta (5)}{64 \pi ^6 R^4} -\sum_{k=1}^{\infty}\frac{e^{-2 \pi k M R} (2 \pi k M R (2 \pi k M R+3)+3) \cos (2 \pi  k Rq_{_b})}{64 \pi ^6 k^5 R^4} \nonumber\\
&=R \left(\frac{\Lambda ^3 M^2}{48 \pi }-\frac{\Lambda M^4}{64 \pi }+\frac{M^5}{60 \pi }\right) +\frac{3 \zeta (5)}{64 \pi ^6 R^4} - \frac{ U(r_b,x)}{128 \pi^6 R^4}
\end{align}
where
\begin{align}\label{Vrx}
	U(r_b,x)\equiv x^2 \text{Li}_3\left(r_b e^{-x}\right)+3 x \text{Li}_4\left(r_b e^{-x}\right)+3 \text{Li}_5\left(r_b e^{-x}\right)+h.c.\,,
\end{align}
with
\begin{align}\label{symbollitium}
	r_b\equiv e^{2\pi i Rq_b} \qquad , \qquad x \equiv 2 \pi R \sqrt{M^2(\phi)},
\end{align}
and $\text{Li}_i(x)$ are the Polylogarithm functions.

If we now calculate the contribution to $V_{1l}(\phi)$ of the corresponding fermion superpartner, we get the same result as in \eqref{bosonfinalp}, with $q_{_b}$ replaced by $q_{_f}$ (and then $r_b$ by $r_f$), and a minus sign overall. Combining these two contributions, the first and second term in the right hand side of \eqref{bosonfinalp} cancel the corresponding ones of the fermion superpartner, and for each couple $(b,f)$ we are left with two finite contributions to $V_{1l}(\phi)$. 
The potential \eqref{EP1} becomes
\begin{align}
\label{final result B+F}
V_{1l}(\phi)=\frac{1}{128 \pi^6 R^4}\sum_{b,f} \left[U(r_f,x)-U(r_b,x)\right].
\end{align}

Let us consider now the models described by \eqref{EP2}, i.e.\,those with $M^2(\phi)=0$ and $q_i=q_i(\phi)$. In this case $V_{1l}(\phi)$ is given by \eqref{bosonfinalp} with the term in parenthesis in the second line vanishing, so that (as observed above) a finite result is obtained separately for bosons and fermions. Irrespectively of being the theory supersymmetric or not, each of the finite boson or fermion contributions to $V_{1l}(\phi)$ is obtained taking $x=0$ (i.e.\,$M^2=0$) and $q_i=q_i(\phi)$ in \eqref{bosonfinalp} ($i=b,f$):
\begin{align}\label{Va}
V^i_{1l}(\phi)=(-1)^{\delta_{fi}}\left(\frac{3 \zeta (5)}{64 \pi ^6 R^4} -\frac{3}{64\pi^6 R^4}\sum_{k=1}^{\infty}\frac{\cos (2 \pi  k Rq_i(\phi))}{k^5}\right).
\end{align}

Eqs.\,\eqref{final result B+F} and \eqref{Va} are the well-known UV-insensitive (finite) results for the two classes of models \eqref{EP1} and \eqref{EP2} respectively (see\cite{Antoniadis:1998sd,Delgado:1998qr} and\cite{Barbieri:2000vh}). For models described by \eqref{EP1} ($M^2 \neq 0$) the strategy of considering the infinite sum over $n$ first is not sufficient to render $V_{1l}(\phi)$ finite:
SUSY is also needed to cancel divergences proportional to powers of $M^2$ (a finite term proportional to $M^5$ is also canceled). For models described by \eqref{EP2} ($M^2=0$) SUSY is unnecessary, and a finite $V_{1l}(\phi)$ is obtained even without supersymmetry. We also stress that within such a calculation strategy the terms that contain $q_i$ (whether they are $q_i$ or $q_i(\phi)$) give rise only to UV-insensitive (finite) contributions, more precisely to oscillatory functions of the $q_i$ (see \eqref{final result B+F} and \eqref{Va}). Interestingly, this oscillatory dependence of $V_{1l}(\phi)$ in \eqref{EP1} and \eqref{EP2} on the $q_i$ is a general property that can be proved even before the calculation is performed. It is due to the fact that there is an infinite sum over $n$, and that the latter is performed independently of the integration over $p$ (in a sense that will be better clarified in the next sections). This property is shown in the Appendix.  

To summarize, in the present section we reviewed the usual approach for the calculation of the one-loop Higgs potential $V_{1l}(\phi)$ from a higher-dimensional theory with compact dimensions. Using the example of a $5D$ theory with one compact dimension, we considered the decomposition of the $5D$ fields in terms of KK-modes, obtained the dimensionally reduced $4D$ action upon integration over the compact dimension $z$, and then calculated $V_{1l}(\phi)$ summing up the Coleman-Weinberg contributions from the infinitely many fields of the KK-towers.

As mentioned in the Introduction, however, this approach needs to be carefully investigated.
In the next section we begin our analysis taking a $5D$ theory with a real scalar field $\widehat \Phi(x,z)$ interacting with a complex scalar field $\widehat \chi(x,z)$, and calculate the $5D$  one-loop potential $\mathcal V^{(5D)}(\widehat \Phi)$. Successively, considering the KK-decomposition of \,$\widehat \Phi(x,z)$ and $\widehat \chi(x,z)$, and choosing $\phi_0(x)$ as the $4D$ Higgs field, we calculate the $4D$ one-loop potential $V_{1l}(\phi_0)$ and establish the relation between $\mathcal V^{(5D)}(\widehat \Phi)$ and $V_{1l}(\phi_0)$. This will allow us to draw important conclusions on the UV-sensitivity of $V_{1l}(\phi_0)$.

\section{Five- and four-dimensional one-loop potential}
Let us consider a $5D$ theory with a real scalar field $\widehat\Phi$ interacting with a complex scalar field $\widehat\chi$, where the fifth dimension is compact in the shape of a circle $S^1$ of radius $R$ (the generalization to phenomenologically viable models is straightforward), with $5D$ (euclidean) action ($a=1,\dots,5$)
\begin{equation}\label{35Daction}
	\mathcal S_{_{(5)}}[\widehat\Phi,\widehat\chi]=\int d^4x \,dz\left(\frac{1}{2}\, \partial_a\widehat\Phi\,\partial^a\widehat\Phi	+ \partial_a\widehat\chi\,\partial^a\widehat\chi^\dagger+\frac{m^2_{\Phi}}{2}\,\widehat\Phi^2+ m^2_\chi\,\widehat\chi \widehat \chi^\dagger+\frac{\widehat\lambda}{4!}\,\widehat\Phi^4+\frac{\widehat g}{2}\,\widehat\Phi^2\widehat\chi \widehat \chi^\dagger\right)\,.
\end{equation}
Taking for $\widehat \Phi(x,z)$ and $\widehat \chi(x,z)$ the boundary conditions (see previous section)
\begin{equation}
	\widehat \Phi(x,z+2\pi R)= \widehat \Phi(x,z) \quad ; \quad \widehat \chi(x,z+2\pi R)= e^{2i\pi R \,q }\,\widehat \chi(x,z)\,,
\end{equation}
we consider their Fourier expansions. As $z$ is compact, and $R$ is much smaller than the size of the other four dimensions, the $5D$ momentum $p^{(5)}$ is ($n$ integer)
\begin{equation}
	p^{(5)} \equiv (p_1,p_2,p_3,p_4,n/R)\equiv (p,n/R)\,.
\end{equation}

To investigate on the UV sensitivity of the dimensionally reduced Higgs one-loop potential $V_{1l}(\phi)$, we begin by considering the $5D$ one-loop potential $\mathcal V^{(5D)}(\widehat \Phi)$
\begin{align}\label{35Dpot}
	&\mathcal V_{1l}^{(5D)}(\widehat \Phi)=\frac{1}{2}{\rm Tr}_{_5}\log\frac{p^2+\frac{n^2}{R^2}+m^2_{\phi}+\frac{\widehat\lambda}{2} \,\widehat \Phi^2}{p^2+\frac{n^2}{R^2}}+
	\frac{1}{2}{\rm Tr}_{_5}
	\log\frac{p^2+\left(\frac{n}{R}+q\right)^2+m^2_\chi+\frac{\widehat g}{2} \,\widehat \Phi^2}{p^2+\frac{n^2}{R^2}} \nn \\
	&=\frac{1}{4\pi R}\sum_n\int\frac{d^4p}{(2\pi)^4}\bigg(\log\frac{p^2+\frac{n^2}{R^2}+m^2_\phi+\frac{\widehat\lambda}{2} \,\widehat \Phi^2}{p^2+\frac{n^2}{R^2}}+\log\frac{p^2+\left(\frac{n}{R}+q\right)^2+m^2_\chi+\frac{\widehat g}{2} \,\widehat \Phi^2}{p^2+\frac{n^2}{R^2}}\bigg)\,,
\end{align}
where the subscript $5$ indicates that the trace is calculated in the $5D$ Fourier space. 

We are already in the position to make some important comments. As
the $5D$ momentum $p^{(5)}=(p,n/R)$ that appears in \eqref{35Dpot} is the momentum of a generic $5D$ virtual particle in the loop, the sum over $n$ and the integral over $p$ are intrinsically intertwined. Therefore in \eqref{35Dpot} we cannot take the asymptotics of one the components of $p^{(5)}$ without considering also the asymptotics of the other components. In particular, we cannot send $n$ to infinity independently\footnote{Note that this is the procedure followed in the usual calculation of the four-dimensional $V_{1l}(\phi)$ (see section 2).} of $p$. These observations show that $\mathcal V^{(5D)}(\widehat \Phi)$ in \eqref{35Dpot} diverges\footnote{If we would perform the infinite sum over $n$ before considering the integral over $p$ (as done in section 2 for the calculation of $V_{1l}(\phi)$), we would erroneously conclude that $\mathcal V^{(5D)}(\widehat \Phi)$ is convergent.}, and that the regularization has to be implemented over the full momentum $p^{(5)}$. From the physical point of view, this simply means that the model has to be regarded as an effective theory, so that the modulus of the $5D$ momentum $p^{(5)}$ should not exceed a maximal UV value above which the theory is no longer valid. In other words, we need to require
\begin{equation}\label{pfive}
	(p^{(5)})^2
	=p^2+n^2/R^2 \leq ( p^{(5)}_{\rm max} )^2 \equiv \Lambda^2\,.
\end{equation} 
This in turn means that in \eqref{35Dpot}, where originally the sum over $n$ and the integral over $p$ are extended up to infinity, we have to make the replacement
\begin{equation}\label{sumint}
\sum_n\int\frac{d^4p}{(2\pi)^5R} \to	\left(\sum_n\int\frac{d^4p}{(2\pi)^5R}\right)'\equiv\frac{1}{2\pi R}\sum_{n=-[R\Lambda]}^{[R\Lambda]} \int^{C^n_{_{\Lambda}}}\frac{d^4p}{(2\pi)^4}\,,
\end{equation}
where
\begin{equation}\label{CnLambda}
	C_{_\Lambda}^n=\sqrt{\Lambda^2-\frac{n^2}{R^2}}
\end{equation}
and $[R \Lambda]$ is the integer part of $R \Lambda$. For the sake of simplicity, in practical calculations we adjust $\Lambda$ in such a way that $R\Lambda$ is an integer.

These considerations are better formulated in a Wilsonian framework, that is at the basis of our physical understanding of quantum field theories. 
Being our $5D$ model not a UV-complete theory but rather an effective one, valid up to a certain UV scale $\Lambda$, the Fourier components of the $5D$ fields $\widehat \Phi(x,z)$ and $\widehat \chi(x,z)$ must be such that $p^{(5)}$ obeys the condition \eqref{pfive}, so that
\begin{align}\label{fouriergiusti}
	\widehat \Phi(x,z)=\left(\sum_n\int\frac{d^4p}{(2\pi)^5R}\right)'\widehat \Phi_{n,p}\,e^{i\left(p\cdot x+\frac{n}{R}z\right)}\,\, ; \,\,\widehat \chi(x,z)=\left(\sum_n\int\frac{d^4p}{(2\pi)^5R}\right)'\widehat \chi_{n,p}\,e^{i\left(p\cdot x+\left(\frac{n}{R}+q\right)z\right)}\,.
\end{align}
Therefore, when calculating the determinant that gives the one-loop potential $\mathcal V^{(5D)}(\widehat \Phi)$, only those eigenvalues of the fluctuation operator that respect this condition have to be taken into account. This means that in \eqref{35Dpot} the replacement \eqref{sumint} has to be made. Finally, from \eqref{fouriergiusti} we have
\begin{align}\label{KKcut}
	\widehat \Phi(x,z)=\frac{1}{\sqrt{2\pi R}}\sum_{n=-R \Lambda}^{R \Lambda} \phi^\Lambda_{n}(x)\,e^{i\frac{n}{R}z}\quad\,\, ; \quad\,\,\widehat \chi(x,z)=\frac{1}{\sqrt{2\pi R}}\sum_{n=-R \Lambda}^{R \Lambda} \chi^\Lambda_{n}(x)\,e^{i\left(\frac{n}{R}+q\right)z}\,,
\end{align}
where $\phi^\Lambda_{n}(x)$ and $\chi^\Lambda_{n}(x)$ are defined through the relations
\begin{equation}\label{KKcutmodes}
	\phi^\Lambda_{n}(x) \equiv \frac{1}{\sqrt{2\pi R}}\int^{C^n_{_\Lambda}}\frac{d^4p}{(2\pi)^4}\,\widehat \Phi_{n,p}\,e^{ip\cdot x} \quad \,\, ; \quad \,\, \chi^\Lambda_{n}(x) \equiv \frac{1}{\sqrt{2\pi R}}\int^{C^n_{_\Lambda}}\frac{d^4p}{(2\pi)^4}\,\widehat \chi_{n,p}\,e^{ip\cdot x}\,.
\end{equation}

Inserting \eqref{KKcut} in \eqref{35Daction}, choosing (similarly to what is done in section 2) $\phi_0^\Lambda(x)$ as the $4D$ Higgs field, and integrating over the compact variable $z$, we end up with a $4D$ action that contains a {\it finite number} of Kaluza-Klein fields $\phi_n^\Lambda(x)$ and $\chi_n^\Lambda(x)$. The calculation of the Higgs one-loop potential for a constant value $\phi$ of $\phi_0^\Lambda(x)$ then gives
\begin{align}\label{pot4d}
	V_{1l}(\phi)=\frac{1}{2}\sum_{n=-R\Lambda}^{R\Lambda}\int^{C^n_{_\Lambda}}\frac{d^4p}{(2\pi)^4}\left(\log\frac{p^2+\frac{n^2}{R^2}+m^2_\phi+\frac{\lambda}{2} \, \phi^2}{p^2+\frac{n^2}{R^2}}+\log\frac{p^2+\left(\frac{n}{R}+q\right)^2+m^2_\chi+\frac{g}{2} \,\phi^2}{p^2+\frac{n^2}{R^2}}\right)\,,
\end{align}
where, in terms of the dimensionful $5D$ couplings $\widehat \lambda$ and $\widehat g$ in \eqref{35Daction} and \eqref{35Dpot}, the $4D$ couplings $\lambda$ and $g$ are defined through the relations
\begin{equation}\label{couplingrd}
	\lambda\equiv\frac{\widehat \lambda}{2\pi R} \quad ; \quad g\equiv\frac{\widehat g}{2\pi R}\,,
\end{equation}
and the combinations
\begin{equation}
	m_{\phi,n}^2\equiv m_\phi^2+\frac{n^2}{R^2} \quad ; \quad m_{\chi,n}^2\equiv m_\chi^2+\left(\frac{n}{R}+q \right)^2
\end{equation}
are the so called KK-masses of each of the $\phi^\Lambda_n(x)$ and $\chi^\Lambda_n(x)$ $4D$ fields.

As a result of the physical cuts in \eqref{KKcut} and \eqref{KKcutmodes}, the $4D$ one-loop Higgs potential in \eqref{pot4d} differs from \eqref{EP1} for the presence of the cuts in the sum over $n$ and integration over $p$. From \eqref{KKcut} and \eqref{KKcutmodes} we also see that the constant value $\widehat \Phi$ of the $5D$ field $\widehat \Phi(x,z)$ in the $5D$ potential \eqref{35Dpot}, and the constant value $\phi$ of the $4D$ field $\phi_0^\Lambda(x)$  in the $4D$ potential \eqref{pot4d} are related, with
\begin{equation}\label{phiPhi}
	\widehat \Phi=\frac{1}{\sqrt{2\pi R}}\, \phi\,,
\end{equation}
and this shows that the $4D$ one-loop Higgs potential $V_{1l}(\phi)$ \eqref{pot4d} (and not \eqref{EP1}) is obtained from the one-loop potential $\mathcal V_{1l}^{(5D)}(\widehat \Phi)$ \eqref{35Dpot} of the original $5D$ theory through the relation 
\begin{equation}\label{rightpot}
	V_{1l}(\phi)=2\pi R \, \mathcal V_{1l}^{(5D)}(\widehat \Phi)\,.
\end{equation}  

Few comments are in order. As already noted below Eq.\,\eqref{35Dpot} for $\mathcal V_{1l}^{(5D)}(\widehat{\Phi})$, the sum over $n$ and the integral over $p$ are intertwined, and we cannot include in the calculation the asymptotics of $n$ independently from the asymptotics of $p$. On the contrary, in Eqs.\,\eqref{EP1} and \eqref{EP2} for the four-dimensional $V_{1l}(\phi)$ it seems that $n$ and $p$ have genuinely different roles: $p$ has the role of the four-momentum of the KK-particles in the $4D$ theory, while $n$ enumerates the {\it infinitely many} KK-fields. From this $4D$ perspective it then seems that to calculate $V_{1l}(\phi)$ we have to sum up the infinitely many Coleman-Weinberg one-loop contributions from the KK-fields. However, from Eq.\,\eqref{rightpot} that gives the connection between $\mathcal V_{1l}^{(5D)}(\widehat \Phi)$ and $V_{1l}(\phi)$, we already know that this amounts to consider the asymptotics of $p_5$ independently of the asymptotics of $p$. This is at the origin of the illusory result that $V_{1l}(\phi)$ is UV-insensitive. In this respect, we also note that the vivid interpretation of $n/R$ as the mass of a KK particle has to be taken with a grain of salt, as it should never be forgotten that $n/R$ is the fifth component of the loop momentum and as such it is deeply intertwined to the other components. It is also worth to note that this latter observation allows to understand that the EFT paradigm is perfectly suited to higher-dimensional theories with compact extra dimensions, and in our opinion overcomes some recently expressed warnings\cite{Burgess:2023pnk}. 

These considerations have an important mathematical counterpart, that helps to shed more lights on the flaws hidden in the usual calculation. Sending $n$ to infinity while keeping the domain of the $p$ integration fixed (that is realized keeping $\Lambda$ fixed), and expanding  only later this domain to cover the entire $\mathbb R^4$ (obtained sending $\Lambda \to \infty$), that is the usual approach to the calculation of $V_{1l}(\phi)$, does not respect the very definition of multi-dimensional improper integrals. To better appreciate this point, it is useful to transform first the infinite sum over $n$ into an integral of a piecewise constant function, thus getting a five-dimensional integral of the kind $\int_{\mathcal D} d^5p f(p^{(5)})$, where $\mathcal D\equiv\mathbb R^5$. To define this improper integral, we must consider a sequence of {\it compact five-dimensional domains} $\mathcal D_i$ (with $\mathcal D_i \mathcal \subset \mathcal D_{i+1}$), that in the limit $i \to \infty$ cover $\mathcal D$. When we integrate (sum) over the fifth component of $p^{(5)}$ in the whole infinite $p_5$ range (infinite $n$ range), that is what is done in the usual approach for the calculation of $V_{1l}(\phi)$ (see section 2), we are rather taking a sequence of {\it non-compact domains} $\mathcal D_i^\infty$, and this can lead (and actually leads) to incorrect results. 
In section 4 we will see that the UV-insensitive result for  $V_{1l}(\phi)$ comes from such an incorrect procedure. When the calculation is performed correctly, UV-sensitive terms appear.

Keeping in mind the above remarks, in the next section  we start our analysis on the UV-sensitivity of $V_{1l}(\phi)$ introducing a hard cutoff (along the lines exposed above) both for the sum over $n$ and for the integration over $p$. In section 5 a similar calculation is performed with the help a smooth cutoff.

\section{UV-sensitivity of $\boldsymbol{V_{1l}(\phi)}$. Hard cutoff}
To begin our analysis on the UV-sensitivity of $V_{1l}(\phi)$, in the present section we perform the calculations introducing a hard cut for the sum over $n$ and the integration over $p$. 
We have already seen in the previous section that as \,$p_1$, $p_2$, $p_3$, $p_4$, $n/R$ \,are the components of the $5D$ loop momentum $p^{(5)}$, the cut on the latter has to be realized requiring (here we write again \eqref{pfive})
\begin{equation}\label{physcut}
(p^{(5)})^2=p^2+\frac{n^2}{R^2} \leq \Lambda^2\,,
\end{equation}
where $p$ was already defined as $p\equiv (p_1,p_2,p_3,p_4)$.

For the purposes of our analysis, however, in section 4.1 we will consider as in\cite{Ghilencea:2001ug} a numerical cut $L$ for the sum over $n$ and a momentum cutoff $\Lambda$ for the integration over $p$, since this allows to get a result for $V_{1l}(\phi)$ that can be compared with those obtained with the usual approach, where the infinite sum over $n$ is performed first. With these two separate cuts, in fact, we can calculate the limit $L\to \infty$ while keeping $\Lambda$ fixed (that corresponds to perform the infinite sum over $n$ first), and compare it with the usual result.
In section 4.2 we will come back to the more physical cut \eqref{physcut}. 

\subsection{Cylindrical hard cutoff}
Let us introduce then a momentum cutoff $\Lambda$ for the integration over $p$, and a numerical cut $L$ for the sum over $n$, that means a cutoff $L/R$ for the fifth component $p_5=n/R$ of the $5D$ momentum   $p^{(5)}= (p_1,p_2,p_3,p_4,p_5)$.
For the sake of definiteness, we begin by considering a single bosonic contribution to $V_{1l}(\phi)$ only.
Moreover, we note that we can treat both cases \eqref{EP1} and \eqref{EP2} at once if we write $V_{1l}(\phi)$ with  the convention that $M^2=M^2(\phi)$ and $q_b$ is field-independent for the case \eqref{EP1}, while $M^2=0$ and $q_b=q_b(\phi)$ for the case \eqref{EP2}. We have:
\begin{equation}
\label{effpotsumcut}
V^b_{1l}(\phi)=\frac{1}{2} \sum_{n=-L}^{L} \int^{\Lambda} \frac{d^4p}{(2\pi)^4}\log \left(\frac{p^2+M^2+(\frac n R+q_b)^2}{p^2+\frac{n^2}{R^2}}\right).
\end{equation}
Performing first the integration over $p$ we obtain (omitting for simplicity the label $b$)
\begin{align}\label{sumL}
&V_{1l}(\phi)=\sum_{n=-L}^{L}\frac{1}{64\pi^2}\Bigg[ \Lambda ^4 \log \frac{\Lambda ^2+M^2+\left(\frac{n}{R}+q\right)^2}{\Lambda ^2+\frac{n^2}{R^2}}+\Lambda ^2 \left(M^2+\left(\frac{n}{R}+q\right)^2-\frac{n^2}{R^2}\right) \nonumber\\ 
&+\left(M^2+\left(\frac{n}{R}+q\right)^2\right)^2 \log \frac{M^2+\left(\frac{n}{R}+q\right)^2}{\Lambda ^2+M^2+\left(\frac{n}{R}+q\right)^2}-\frac{n^4}{R^4}\log \frac{\frac{n^2}{R^2}}{ \Lambda^2+\frac{n^2}{R^2}} \Bigg] \equiv \sum_{n=-L}^{L}F(n).
\end{align}

An important novelty of our analysis is that we calculate \eqref{sumL} with the help of the Euler-McLaurin (EML) formula. This turns out to be a good choice since, differently from previous attempts\cite{Ghilencea:2001bv,Ghilencea:2001ug,Kim:2001gk}, it allows us to obtain for the first time the complete result for $V_{1l}(\phi)$. We then write
\begin{align}\label{EML}
V_{1l}(\phi)&=\int_{-L}^L dx\, F(x)+\frac{F(L)+F(-L)}{2}+\sum_{k=1}^{r}\frac{B_{2k}}{(2k)!}\left(F^{(2k-1)}(L)-F^{(2k-1)}(-L)\right)+R_{2r},
\end{align} 
where $r$ is an integer, $B_n$ are the Bernoulli numbers, and the rest $R_{2r}$ is given by
\begin{equation}\label{resto}
R_{2r}=\sum_{k= r+1}^{\infty}\frac{B_{2k}}{(2k)!}\left(F^{(2k-1)}(L)-F^{(2k-1)}(-L)\right)=\frac{(-1)^{2r+1}}{(2r)!}\int_{-L}^{L}dx\,F^{(2r)}(x)B_{2r}(x-[x]),
\end{equation}
with $B_n(x)$ the Bernoulli polynomials, while $[x]$ indicates the integer part of $x$. 

Clearly, if in  \eqref{sumL}, \eqref{EML} and \eqref{resto} we send $L \to \infty$ while keeping $\Lambda$ fixed, we get for $V_{1l}(\phi)$ the celebrated UV-insensitive (finite) result of section 2. From the discussion of section 3, however, we know that the asymptotics of the different components of $p^{(5)}$ cannot be included in the calculation of $V_{1l}(\phi)$ independently of one another. As a consequence, we cannot take the $L \to \infty$ limit while keeping $\Lambda$ fixed. In the process of including the asymptotics of the loop momentum in $V_{1l}(\phi)$, on the contrary, we have to keep the ratio $L/R \Lambda$ finite, that is we cannot introduce any hierarchical order in the formal limits $L,\Lambda \to \infty$. In other words the asymptotics are properly taken into account if we include them in  \eqref{effpotsumcut} while keeping the ratio $L/R\Lambda$ finite:
\begin{equation}\label{finlimit}
	\frac{L}{R\Lambda} \,\,\,\,\,\, {\rm finite\,\,\,\,\,\,when}\,\,\,\,\,\,L,\Lambda \to \infty\,.
\end{equation}
Moreover, from the physical meaning of the UV cuts, we know that the right hand side of \eqref{effpotsumcut} has to be considered only for values of $M$ and $q$ that fulfill the conditions
\begin{equation}\label{limitsUV}
	M^2,q^2 \ll \Lambda^2, L^2/R^2\,.
\end{equation} 
The conditions \eqref{finlimit} and \eqref{limitsUV} are easily implemented in our calculations if we write ($\xi$ dimensionless finite number)
\begin{equation}\label{replacement}
L= \xi R \Lambda\,,
\end{equation}
and expand each term in \eqref{EML} for \,${M^2}/{\Lambda^2},{q^2}/{\Lambda^2} \ll 1$. 

After inserting \eqref{replacement} in \eqref{EML}, we see that the first term in the sum over $k$ in the latter equation (i.e.\,\,the term with $k=1$) turns out to be linearly divergent ($\sim \Lambda$), while those starting from $k=2$ are suppressed by inverse powers of $\Lambda$, more precisely they behave as $\Lambda^{3-2k}$. In particular, as the term with $k=2$ is $\mathcal{O}(\Lambda^{-1})$, for any $r>1$ we have  $R_{2r}=R_2+\mathcal{O}(\Lambda^{-1})$. Therefore, we can (and will) perform the calculation by referring always to the rest $R_2$.
Expanding \eqref{EML} for\, ${M^2}/{\Lambda^2},\,{q^2}/{\Lambda^2} \ll 1$, we finally obtain ($\xi$ is finite)
\begin{align}\label{divpot}
&V_{1l}(\phi)=\frac{ 2 M^2 \tan ^{-1}\xi +\xi  \left(\xi ^2 \log \frac{\xi ^2}{\xi ^2+1}+1\right) \left(M^2+3 q^2\right)}{48 \pi ^2} \,R \Lambda ^3\nonumber\\
&+\frac{\xi ^2 \left(M^2+3 q^2\right)+\xi ^2 \left(\xi ^2+1\right)\left(M^2+3 q^2\right) \log \frac{\xi ^2}{\xi ^2+1} +M^2+q^2}{32 \pi ^2 \left(\xi ^2+1\right)}\,\Lambda ^2\nonumber\\
&+\frac{\xi \, M^2 \left(6 q^2 R^2+1\right)\left(\xi ^2+1\right)+\xi \,q^2 \left(q^2 R^2+1\right)\left(3 \xi ^2+5\right) }{96 \pi ^2 \left(\xi ^2+1\right)^2}\,\frac{\Lambda  }{R}\nonumber\\
&+ \frac{\xi  \log \frac{\xi ^2}{\xi ^2+1} \left(3 R^2 \left(M^2+q^2\right)^2+M^2+3 q^2\right)-3 M^4 R^2 \tan ^{-1}\xi }{96 \pi ^2}\,\frac{\Lambda  }{R}\nonumber
\end{align}
\begin{align}
&+ \frac{3 \left(\xi ^2+1\right)^2 M^4+6 \left(\xi ^4+4 \xi ^2+3\right) M^2 q^2+\left(3 \xi ^4+6 \xi ^2+11\right) q^4}{192 \pi ^2 \left(\xi ^2+1\right)^3}\nonumber\\
&+\frac{16 \pi  M^5 R+15 \log \frac{\xi ^2}{\xi ^2+1} \left(M^2+q^2\right)^2}{960 \pi ^2} +R_2+\mathcal{O}\left(\Lambda^{-1}\right).
\end{align}

To realize the previously mentioned comparison of our result \eqref{divpot} with the usual calculations present in the literature (section 2), we now consider its limit for $\xi \to \infty$, with $\Lambda$ kept finite. 
We get\footnote{Resorting directly to \eqref{EML}, and taking the $L\to\infty$ limit (while keeping $\Lambda$ finite), for $V_{1l}(\phi)$ we get (see \eqref{divm})
	\begin{align*}
		V^{1l}(\phi)= -\frac{\Lambda ^5 R}{40 \pi }+\frac{M^5 R}{60 \pi }+\frac{\Lambda ^2 M^2 R \sqrt{\Lambda ^2+M^2}}{120 \pi }+\frac{\Lambda ^4 R \sqrt{\Lambda ^2+M^2}}{40 \pi }-\frac{M^4 R \sqrt{\Lambda ^2+M^2}}{60 \pi }+R_2\,.
	\end{align*}
	This coincides with \eqref{zetalimit2} if we expand for $M^2/\Lambda^2\ll1$.} ($\widetilde{R}_2\equiv\lim_{\xi \to \infty} R_2$)
\begin{equation}\label{zetalimit2}
V_{1l}(\phi)\widesim[1.75]{\xi\gg 1} \frac{R\Lambda ^3 M^2}{48 \pi }-\frac{R\Lambda M^4}{64 \pi }+\frac{RM^5}{60 \pi }+\widetilde R_2+\mathcal O\left(\xi^{-1}\right).
\end{equation} 
Due to \eqref{replacement}, in this limit we must obtain the result \eqref{bosonfinalp} of section 2.
The comparison of \eqref{zetalimit2} with \eqref{bosonfinalp} then shows that, in the $\xi \to \infty$ limit, $R_2$ has to coincide with the sum of the second and third term in the right hand side of \eqref{bosonfinalp} (see also \eqref{Vrx} and \eqref{symbollitium}). Restoring the label $b$, we finally have 
\begin{align}\label{R2}
\widetilde{R}_2=\lim_{\xi \to \infty} R_2=\frac{3 \zeta (5)}{64 \pi ^6 R^4}-\frac{1}{128 \pi^6 R^4} \left[x^2 \text{Li}_3\left(r_b e^{-x}\right)+3 x \text{Li}_4\left(r_b e^{-x}\right)+3 \text{Li}_5\left(r_b e^{-x}\right)+h.c.\right]\,.
\end{align}
The corresponding fermion contribution has the same form with a minus sign overall and $r_b$ replaced by $r_f$.  

To make the correct calculation with no hierarchy between the asymptotics of $L$ and $\Lambda$, we need to evaluate $R_2$ for finite $\mathcal O(1)$ values of $\xi$. 
To this end we resort to \eqref{resto}, which provides an explicit formula to calculate $R_2$. Although we cannot perform this calculation analytically, using for the Bernoulli polynomials $B_{2r}(x-[x])$ the relation
\begin{equation}\label{B}
	B_{2r}(x-[x])=(-1)^{r-1}(2r)!\sum_{k=1}^\infty \frac{2 \cos(2\pi k x)}{(2\pi k)^{2r}}\,,
\end{equation}
we can resort to a numerical analysis.

\begin{figure}[t]
	\centering
	\includegraphics[scale=0.2]{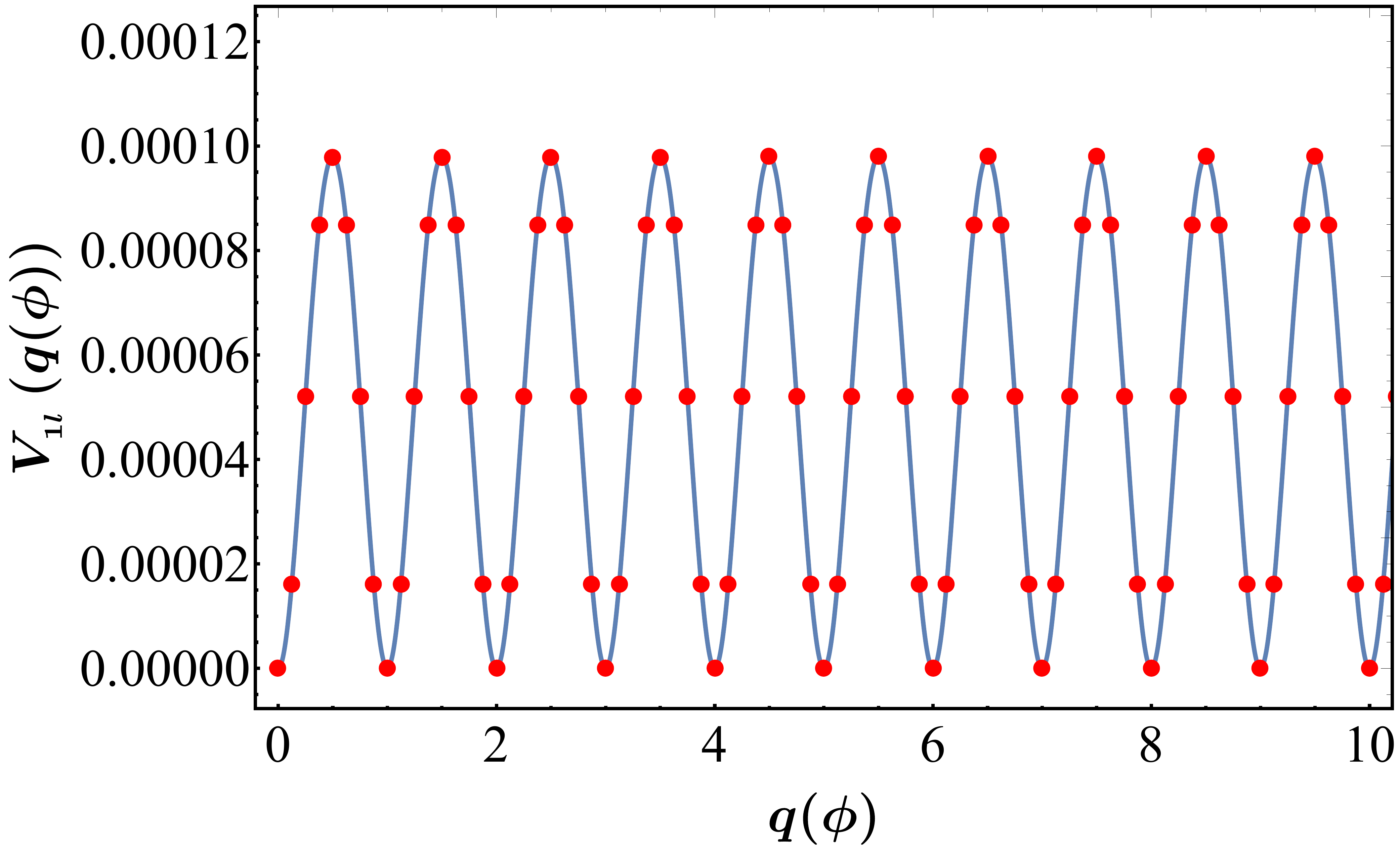}
	\hspace{3mm}
	\includegraphics[scale=0.2]{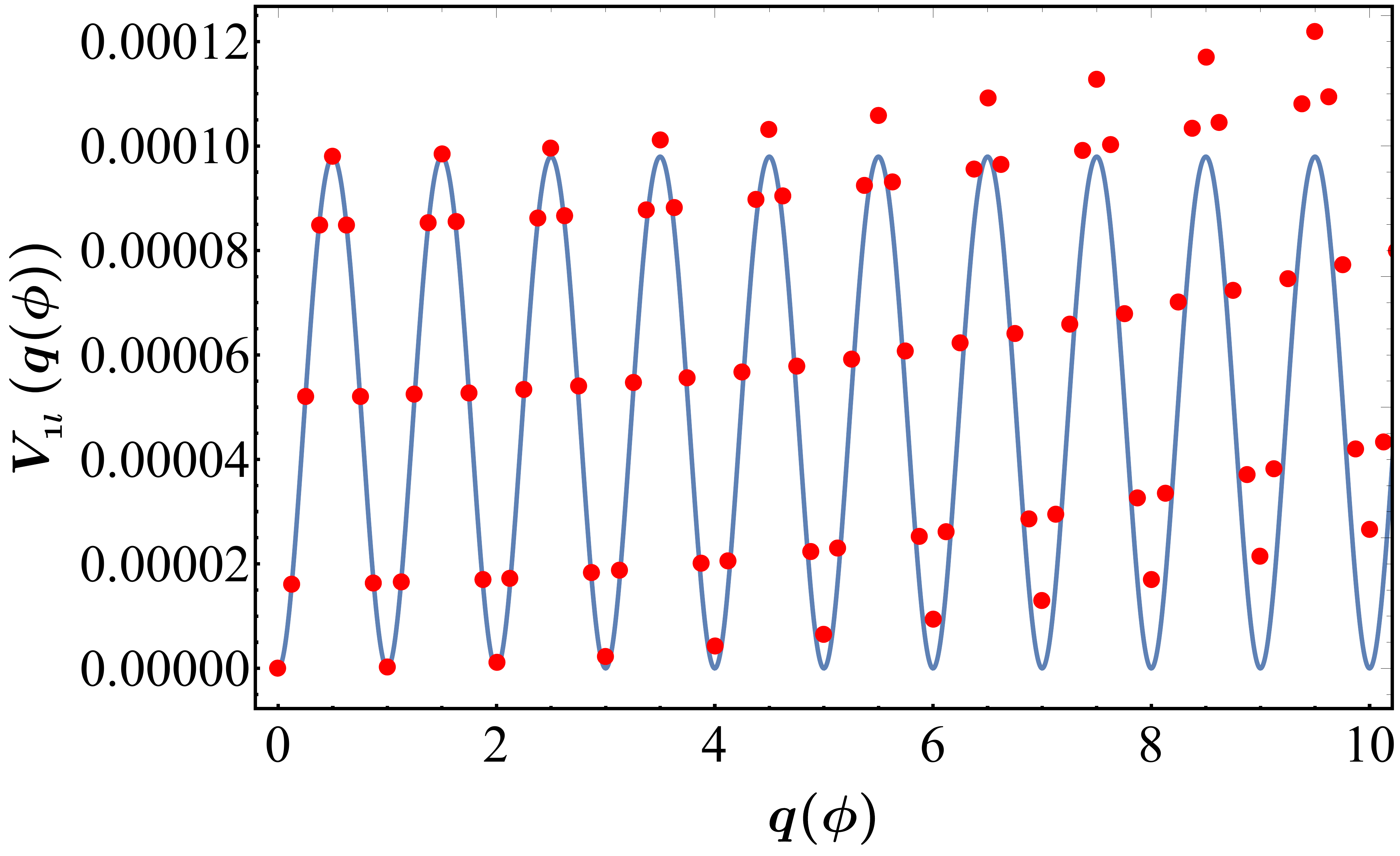}
	\caption{{\it Left panel}: the rest $R_{2r}$ for $r=1$ as a function of $q$ in the range $q \in [0,10]$, for $\xi=1$, $R=1$, $\Lambda=5\times10^4$ and $M=0$. The red dots come from the numerical integration of \eqref{resto}. The continuous blue curve is $R_2(q)$ for the same values of $R$ and $M$ obtained for $\xi \to \infty$, Eq.\,\eqref{R2}. {\it Right panel}: the same as in the left panel, with $\Lambda=10^2$.}
	\label{Interpolation}
\end{figure}

It is convenient to consider separately the two cases \eqref{EP1} and \eqref{EP2}.
Let us begin with \eqref{EP2}, i.e.\,with the case $M=0$ and $q=q(\phi)$. Taking $\xi=1$, and assigning also specific values to $R$ and $\Lambda$, with the help of \eqref{resto} we evaluate numerically $R_2$ as a function of $q$. In the left panel of Fig.\,\ref{Interpolation}, the result for $R=1$ and $\Lambda=5\times10^4$ in the range $q \in [0,10]$ is given by the red dots, while the continuous blue line is the plot of $\widetilde R_2$ in \eqref{R2} for the same value of $R$. 
We see that, under the condition $q^2/\Lambda^2 \ll 1$, even when $\xi \sim 1$, the rest $R_2$ coincides with $\widetilde{R}_2$ (in this case $M=0$, i.e.\,\,$x=0$). As we know, \eqref{R2} is nothing but (a bosonic contribution to) the celebrated UV-insensitive one-loop Higgs potential \eqref{Va} of the usual approach. We have then found that such a finite contribution is present also for values of $\xi \sim 1$ as long as the necessary condition $q^2(\phi)/\Lambda^2 \ll 1$ is satisfied.

It is worth to stress the centrality of this latter point. To this end we use the example of a four-dimensional  $\phi^4$ theory, for which, after performing the loop integral $\frac{\hbar}{2}\int^{(\Lambda)}\frac{d^4k}{(2\pi)^4}
\ln(1+\frac{m_0^2 + \frac{\lambda_0}{2}\phi^2}{k^2})$, the one-loop potential reads\footnote{In $M^2(\phi)$ the bare parameters $m^2_0$ and $\lambda_0$ are replaced with the renormalized ones $m^2$ and $\lambda$ as the counterterms are $\mathcal O(\hbar)$ and the loop correction already contains the $\hbar$ factor.}
\begin{eqnarray}\label{effpotdimreg-c}
	U\left(\phi\right)
= \frac{m_0^2}{2}\phi^2
	+\frac{\lambda_0}{4!}\phi^4
	+\frac{\hbar}{64\pi^2}\Biggl[\Lambda^4\,{\ln}
	\left(1+\frac{M^2(\phi)}{\Lambda^2}
	\right)
	+M^2(\phi)\,\Lambda^2  
	-\left(M^2(\phi)\right)^2 
	{\ln}
	\left(\frac{\Lambda^2}
	{M^2(\phi)}+1\right)\Biggr],
\end{eqnarray}
where $M^2(\phi)=m^2+\frac \lambda 2 \phi^2$.
As it is well known,  \eqref{effpotdimreg-c} is {\it physically meaningful} only for $M^2(\phi)/\Lambda^2 \ll 1$. Expanding \eqref{effpotdimreg-c} in powers of $M^2(\phi)/\Lambda^2$
\begin{align}\label{UVscalarpot}
U (\phi)=\frac{m_0^2}{2}\phi^2+\frac{\lambda_0}{4!}\phi^4
	+\frac{\hbar}{64\pi^2}\bigg[2\Lambda^2\,M^2(\phi) +\left(M^2(\phi)\right)^2 
	\left(\ln\frac{M^2(\phi)}{\Lambda^2}
	-\frac{1}{2}\right)
	\bigg], 
\end{align}
that finally becomes the renormalized potential once the bare parameters are written in terms of the renormalized ones and of the counterterms (that eventually cancel the divergences).

This example underlines the importance of the condition $q^2(\phi)/\Lambda^2 \ll 1$ in our case. This is even better appreciated if we consider again Eq.\,\eqref{resto} for $R_2(q)$ in the range $q \in [0,10]$ with $R=1$ (as for the left panel of Fig.\,\ref{Interpolation}), but with a lower cutoff, $\Lambda=10^2$. The result of the numerical analysis is given by the red dots in the right panel of Fig.\,\ref{Interpolation}. As for the left panel, the blue continuous line is the plot of \eqref{R2} for the same value of $R$. Comparing with the left panel, we see that the superposition between dotted and continuous curve  holds for a more limited range of $q$, since in this case the condition $q^2/\Lambda^2 \ll 1$ breaks up for smaller values of $q$.

The important result that emerges from this analysis is that the finite term \eqref{R2}, that is the only outcome of the calculation for $V_{1l}(\phi)$ when the usual approach is considered, does not come from the fact that the infinite sum over $n$ is performed, but rather   from the physical requirement $q^2/\Lambda^2 \ll 1$, even when the limits $L \to \infty$, $\Lambda \to \infty$ are properly considered, i.e.\,when the ratio $L/R \Lambda$ is kept finite ($\xi$ finite). We will further comment on this point later. For the time being we observe that in the present case ($M=0$ and $q$ field dependent), from \eqref{divpot} we have for $V_{1l}(\phi)$
\begin{align}\label{divpot2}
	V_{1l}(\phi)&=\frac{3\xi  \left(\xi ^2 \log \frac{\xi ^2}{\xi ^2+1}+1\right)}{48 \pi ^2}\,q^2(\phi) \,R \Lambda ^3
	+\frac{3 \xi ^2 +3\xi ^2 \left(\xi ^2+1\right) \log \frac{\xi ^2}{\xi ^2+1}+1}{32 \pi ^2 \left(\xi ^2+1\right)}\,q^2(\phi)\,\Lambda ^2\nonumber\\
	&+\frac{\xi \left(3 \xi ^2+5\right)+3\xi \left(\xi ^2+1\right)^2 \log \frac{\xi ^2}{\xi ^2+1} }{96 \pi ^2 \left(\xi ^2+1\right)^2}\,\left( R^2q^4(\phi)+q^2(\phi)\right)\,\frac{\Lambda}{R}
	\nonumber \\
	&+ \left(\frac{3 \xi ^4+6 \xi ^2+11}{192 \pi ^2 \left(\xi ^2+1\right)^3}+\frac{\log \frac{\xi ^2}{\xi ^2+1} }{64 \pi ^2}\right)q^4(\phi) +\widetilde R_2+ \mathcal{O}\left(\Lambda^{-1}\right)\,,
\end{align}
where we see that, in addition to $\widetilde{R}_2$, that is nothing but the result obtained with the usual calculation, UV-sensitive field-dependent terms (cubic, quadratic and linear divergences in $\Lambda$), and a new finite term, contribute to $V_{1l}(\phi)$. From \eqref{divpot2} we see that in the limit $\xi \to \infty$ only the term $\widetilde R_2$ survives, i.e.\,(as expected) we recover the usual finite result.
\begin{figure}[t]
	\centering
	\includegraphics[scale=0.2]{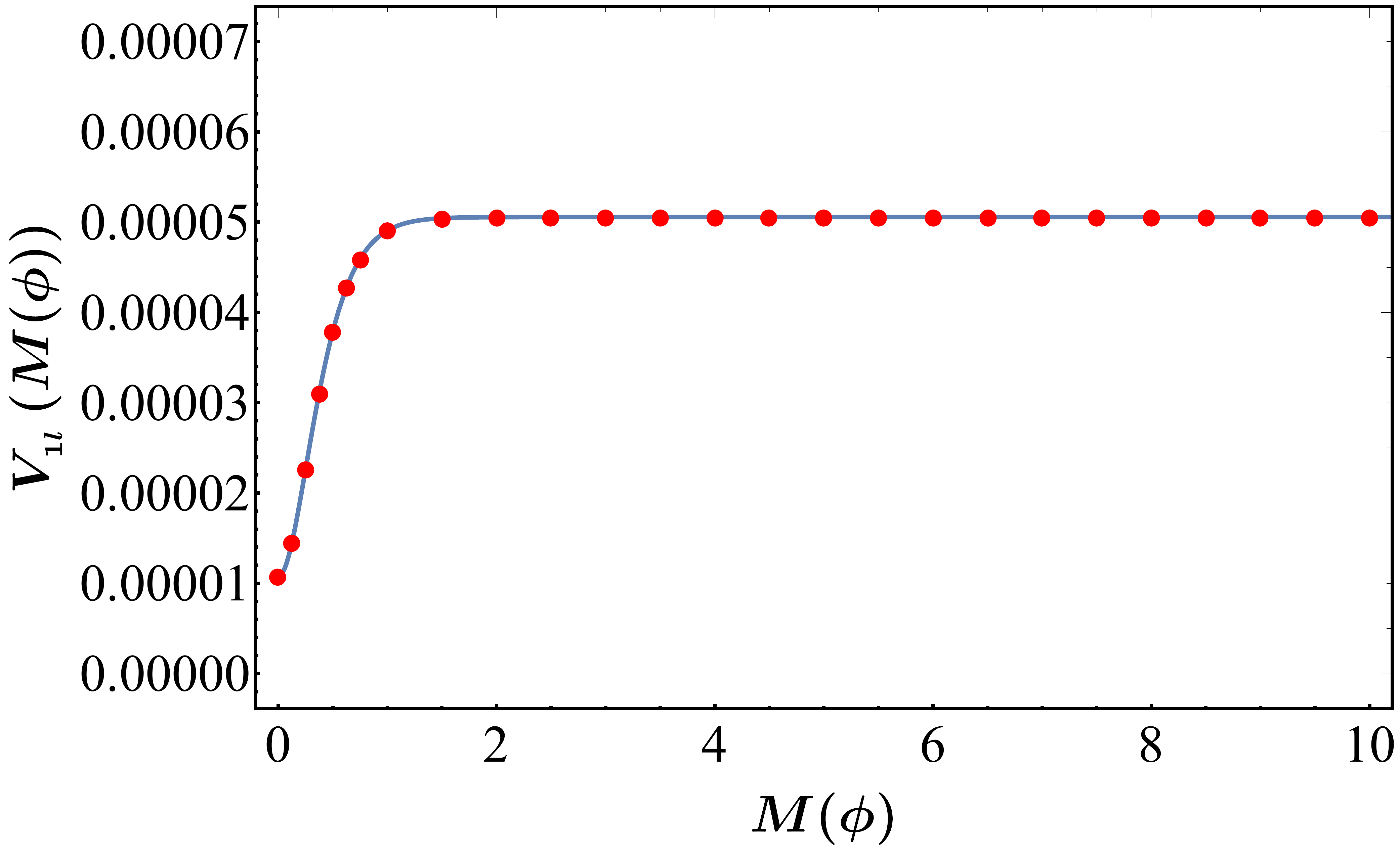}
	\hspace{3mm}
	\includegraphics[scale=0.2]{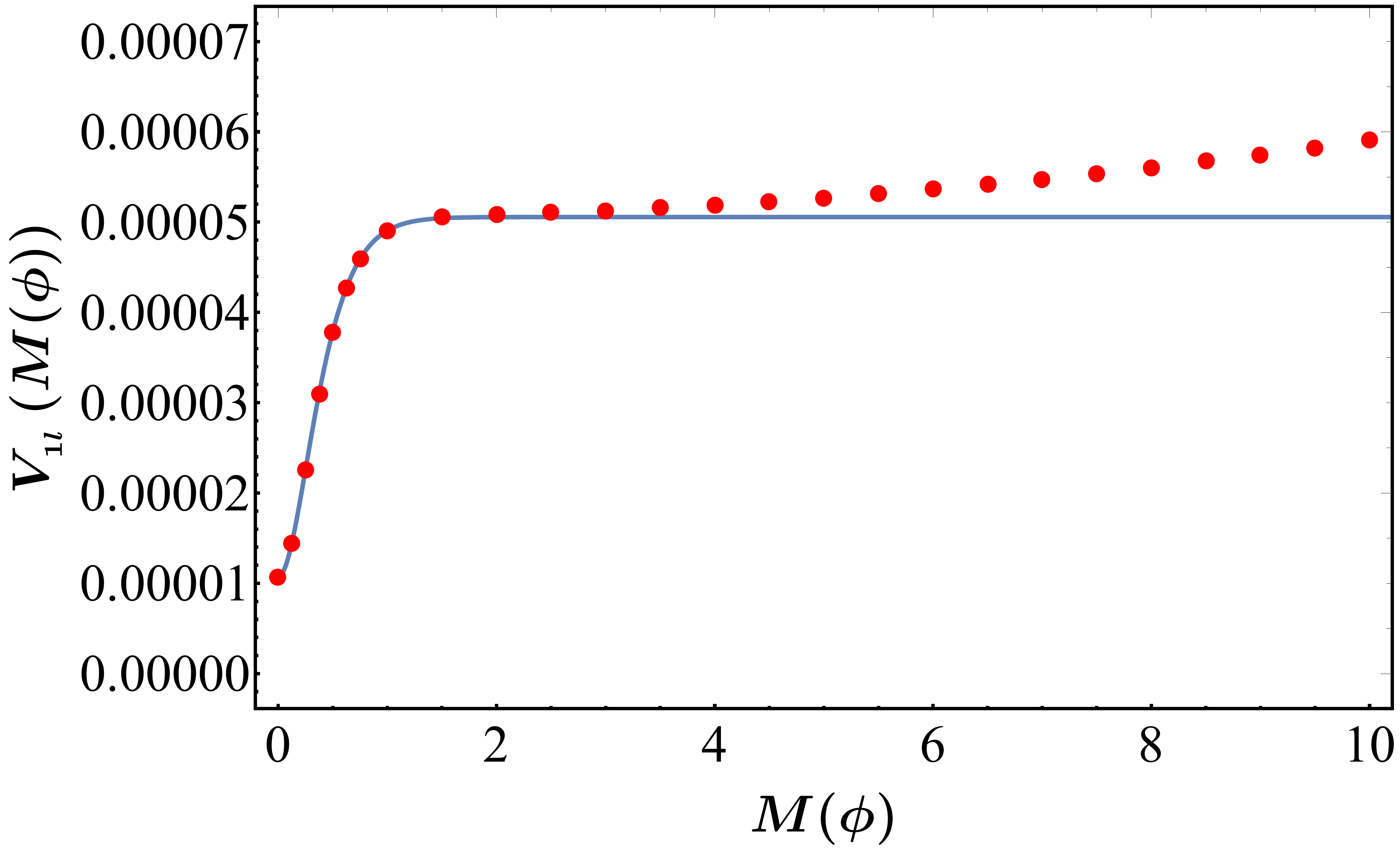}
	\caption{{\it Left panel}: the rest $R_{2r}$ for $r=1$ as a function of $M$ in the range $M \in [0,10]$, for $\xi=1$, $R=1$, $\Lambda=5\times10^4$ and $q=0.1$. The red dots come from the numerical integration of \eqref{resto}. The continuous blue curve is $R_2(M)$ for the same values of $R$ and $q$ obtained for $\xi \to \infty$, Eq.\,\eqref{R2}. {\it Right panel}: the same as in the left panel, with $\Lambda=10^2$.}
	\label{Interpolationm}
\end{figure}	

Let us consider now $V_{1l}(\phi)$ in \eqref{EP1}, i.e.\,\,for models with $M^2=M^2(\phi)$ and $q$ field-independent.
As for the previous case, we take $\xi=1$, assign specific values to $R$, $q$ and $\Lambda$, and evaluate numerically the rest $R_2$ in \eqref{resto}, this time as a function of $M$. In the left panel of Fig.\,\ref{Interpolationm}, $R_2(M)$ is given by the red dots, with $R=1$, $q=0.1$, $\Lambda=5\times10^4$ in the range $M \in [0,10]$. The continuous blue line is the plot of $\widetilde{R}_2(M)$ given in \eqref{R2} for the same values of $R$ and $q$. 
Similarly to the previous case, we see that, even when $\xi \sim 1$,  the rest $R_2$ 
in \eqref{resto} coincides with $\widetilde{R}_2$ in \eqref{R2} when the conditions $M^2(\phi)/\Lambda^2,q^2/\Lambda^2 \ll 1$ are fulfilled. 

To better appreciate the importance of these physical conditions, in the right panel of Fig.\,\ref{Interpolationm} we plot again \eqref{resto} for $R_2(M)$ in the range $M \in [0,10]$, where (as in the left panel) we take $R=1$ and $q=0.1$, but consider  now a lower cutoff $\Lambda=10^2$. The result is given by the red dots. As for the left panel, the blue continuous curve is the plot of \eqref{R2} for the same values of $R$ and $q$, but differently from that case, the superposition between dotted  and  continuous curve holds for a more limited range of $M$, since  in this case the condition $M^2/\Lambda^2 \ll 1$ breaks up for smaller values of $M$.

Let us consider in particular a SUSY model. From \eqref{divpot} we see that the contribution to $V_{1l}(\phi)$ from a single couple of superpartners with charges $q_b$ and $q_f$ is 
\begin{align}
	\label{divpotsusy}
	V_{1l}(\phi)&=\frac{q_b^2-q_f^2}{{16 \pi ^2}}  M^2(\phi) \left(2 \xi\log \frac{\xi ^2}{\xi ^2+1}  +\frac{2\xi}{\xi ^2+1}\right) \Lambda  R \nonumber \\ 
	&+\frac{q_b^2-q_f^2}{{16 \pi ^2}}  M^2(\phi) \left(\log \frac{\xi ^2}{\xi ^2+1} +\frac{\xi ^2+3}{\left(\xi ^2+1\right)^2}\right)+\widetilde R_2^{(b)}-\widetilde R_2^{(f)}+\mathcal{O}\left(\Lambda^{-1}\right),
\end{align}
and that, as long as $\xi$ is kept finite, together with the usual result (that in this case consists of the difference $\widetilde R^{(b)}_2-\widetilde R_2^{(f)}$), an additional finite term and an additional $\Lambda$-sensitive term (linear divergence) are present. It is immediate to see from \eqref{divpotsusy} that, when $\xi \to \infty$, these terms disappear and we are left with the well-known finite result of section 2. Moreover, from  \eqref{divpot} and \eqref{divpotsusy} we see that, while the UV-sensitive terms proportional to powers of $M^2(\phi)$ (but that do not depend on the $q_i$) are cancelled by the presence of superpartners, this does not happen for the terms $q_i^2 M^2(\phi) R \Lambda$, with the exception of the  trivial case $q_b=q_f$.

An important outcome of the present analysis is that in  \eqref{divpot2} and \eqref{divpotsusy} UV-sensitive terms proportional to powers of the $q_i$ appear. They are lost when the calculation is done in the usual manner (section 2), that in this section is realized with the $\xi \to \infty$ limit, but when we take into account the asymptotics of the loop momenta properly, we find that $V_{1l}(\phi)$ is UV-sensitive. Further important comments on the physical meaning and origin of such an UV-sensitivity will be done at the end of the next section, where a spherical cutoff is used.

\subsection{Spherical hard cutoff}
In the present section we consider the spherical cutoff $p^2+p_5^2 \leq \Lambda^2$ introduced in \eqref{physcut}, that immediately leads to the cutoff $C_{_\Lambda}^n=\sqrt{\Lambda^2-n^2/R^2}$ in \eqref{CnLambda} for the integration over $p$, while for the sum over the integer $n$ it is $n=-[R \Lambda], \dots, [R \Lambda]$ (with $[R \Lambda]$ integer part of $R \Lambda$). Assuming for the sake of simplicity that $\Lambda$ is adjusted so that $R\Lambda$ is an integer ($R$ is fixed), for $V_{1l}(\phi)$ we have 
\begin{equation}\label{effpotsumcuti}
	V_{1l}(\phi)=\frac{1}{2} \sum_{n=-R \Lambda}^{R \Lambda} \int^{C^n_{_\Lambda}} \frac{d^4p}{(2\pi)^4}\log \left(\frac{p^2+M^2+(\frac n R+q)^2}{p^2+\frac{n^2}{R^2}}\right)\,,
\end{equation}
and performing the integration over $p$:
\begin{align}\label{sumiso}
V_{1l}(\phi)&=\sum_{n=-R \Lambda}^{R \Lambda}\frac{1}{64\pi^2}\Bigg[ \left(\Lambda ^2-\frac{n^2}{R^2}\right) \left(M^2-\frac{n^2}{R^2}+\left(\frac{n}{R}+q\right)^2\right) -\frac{n^4}{R^4}\log \frac{n^2}{\Lambda ^2 R^2}\nonumber\\
&+\left(M^2+\left(\frac{n}{R}+q\right)^2\right)^2 \log \frac{M^2+\left(\frac{n}{R}+q\right)^2}{\Lambda ^2+M^2-\frac{n^2}{R^2}+\left(\frac{n}{R}+q\right)^2}\nonumber\\
&+\left(\Lambda ^2-\frac{n^2}{R^2}\right)^2 \log \frac{\Lambda ^2+M^2-\frac{n^2}{R^2}+\left(\frac{n}{R}+q\right)^2}{\Lambda ^2}\Bigg].
\end{align}
As in section 4.1, we calculate the sum over $n$ with the help of the EML formula \eqref{EML} and truncate the sum over $k$ at $r=1$. We obtain
\begin{align}\label{divpoti}
&V_{1l}(\phi)=\frac{5 M^2+3 q^2}{180 \pi ^2}\,R\Lambda^3 -\frac{35 M^4+14 M^2 q^2+3 q^4}{840 \pi ^2}\, R \Lambda +\frac{M^5 R}{60 \pi }+R_2+\mathcal{O}(\Lambda^{-1}).
\end{align}
Contrary to what happens with the cylindrical regularization of section 4.1, from \eqref{sumiso} we see that the cubic and linear terms in $\Lambda$ come from the first term in the right hand side of the EML formula \eqref{EML}, i.e.\,\,from the integral, while the second and third term vanish. This means that, with the exception of $R_2$, all the terms in \eqref{divpoti} come from the integral. Therefore, the difference between having a sum (as in the present case) rather than an integral over $p_5$ is entirely encoded\footnote{In the cylindrical regularization, where $p$ and $p_5$ are treated asymmetrically, we see the appearance of a term proportional to $\Lambda^2$ from the second term of the {EML} formula, and a term proportional to $\Lambda$ from the third one. However, both in the cylindrical and spherical regularization, the $k=2$ term in the sum over $k$ is $\mathcal{O}(\Lambda^{-1})$.} in the presence of the 
rest $R_2$.

Eq.\,\eqref{divpoti} looks similar to \eqref{divpot} (where, however, an additional term proportional to $\Lambda^2$ and an additional finite term are also present), but the terms proportional to $\Lambda^3$ and $\Lambda$ have different numerical coefficients.
This mismatch is easily understood if we note that treating the cut in the $5D$ momentum in a symmetric manner (as done in this section) and introducing two separate cuts for the modulus of the first four components $p=\sqrt{p_1^2+p_2^2+p_3^2+p_4^2}$ and for the fifth component $p_5$ (as done in section 4.1) are clearly two different operations. 
In this latter case, in fact, the integration over $p$ reproduces the typical results of a $4D$ theory, including the quadratic divergences. Obviously, performing the integration over $p^{(5)}$ (that for the fifth component $n/R$ means $\sum_n$) with a spherical cutoff, cannot generate any term proportional to $\Lambda^2$. By the same token, the additional finite terms in \eqref{divpot} are again due to the asymmetric way of treating the cuts in $n/R$ and in $p$.

\begin{figure}[t]
	\centering
	\includegraphics[scale=0.2]{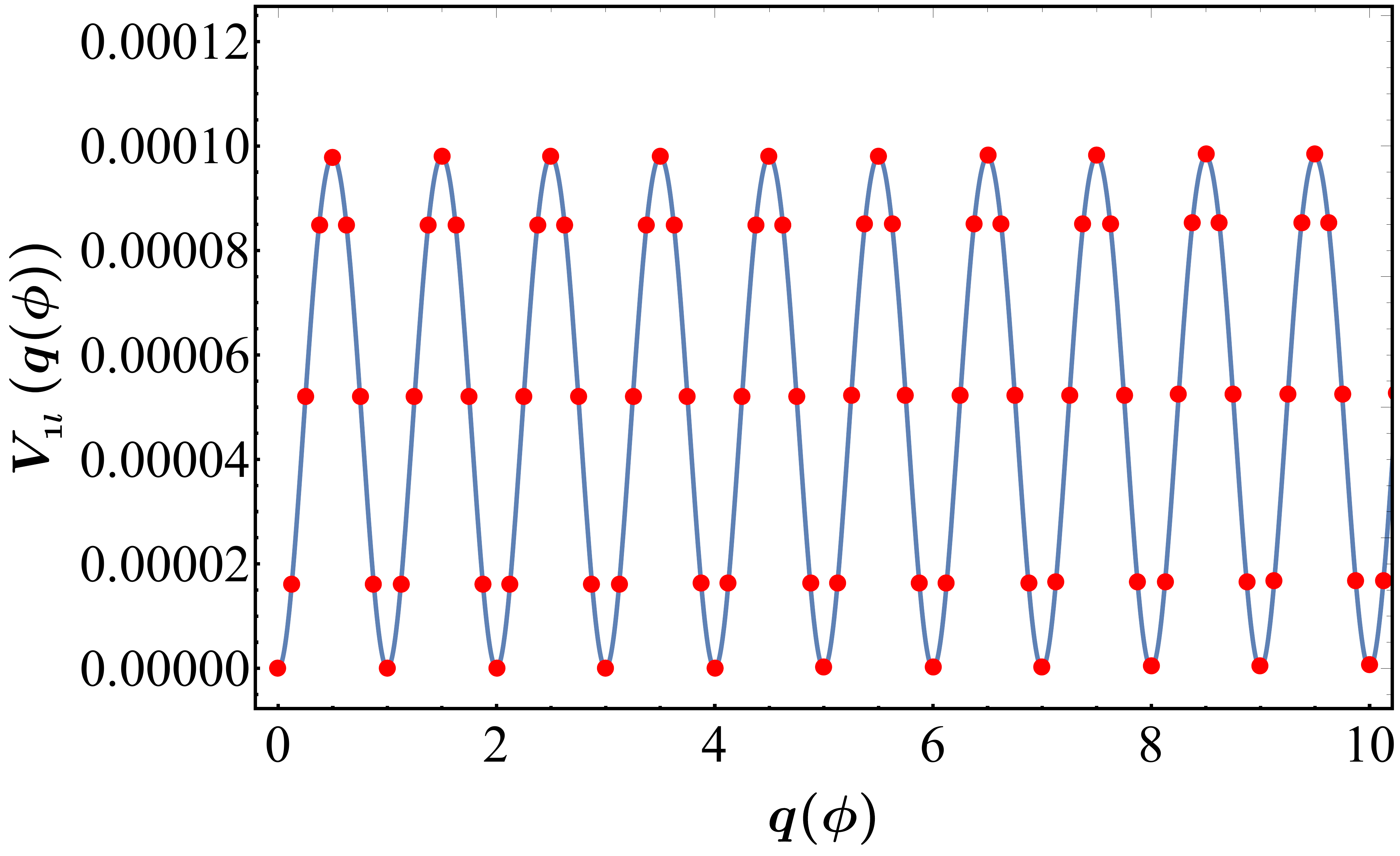}
	\hspace{3mm}
	\includegraphics[scale=0.2]{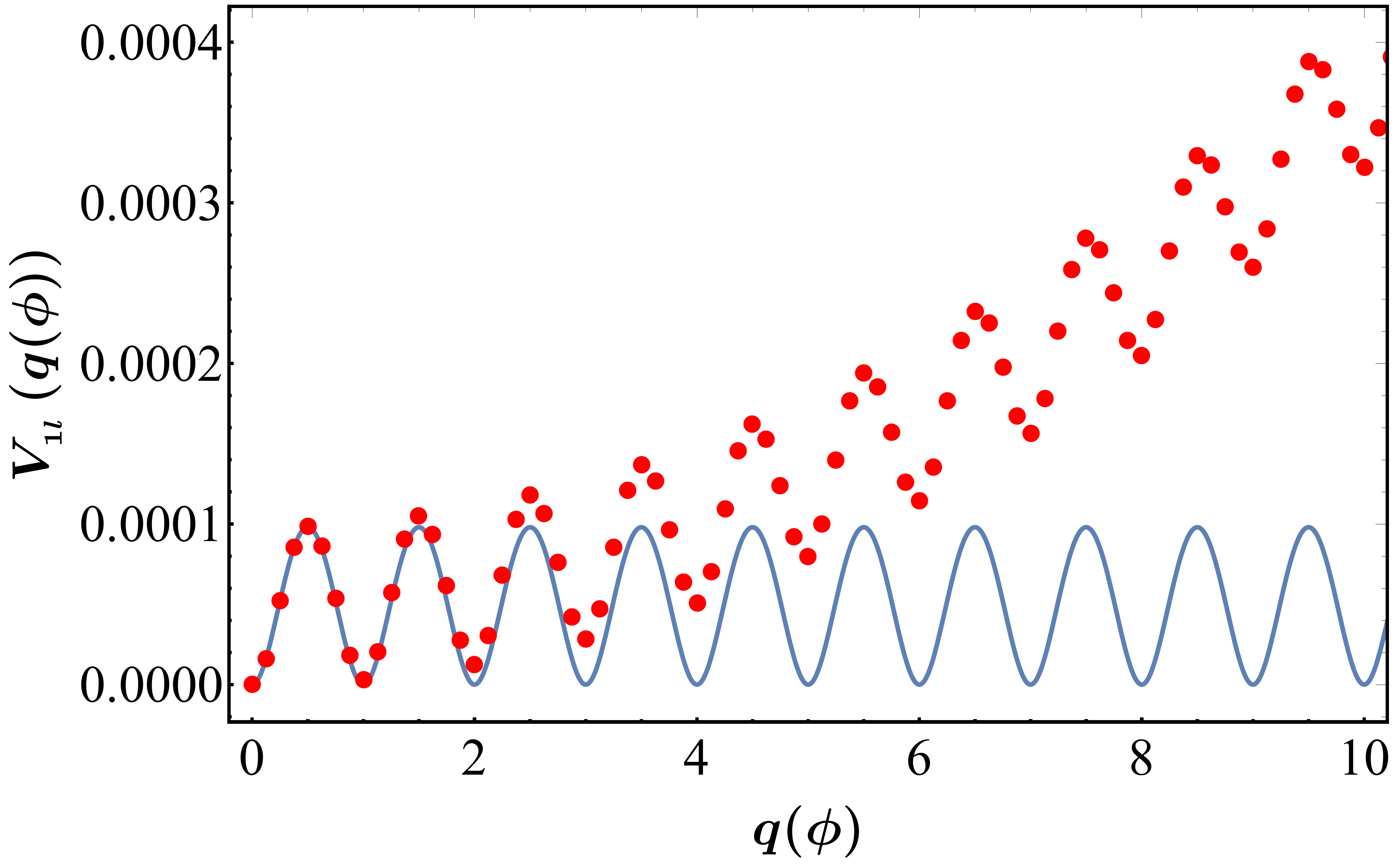}
	\caption{{\it Left panel}: the rest $R_{2r}$ for $r=1$ as a function of $q$ in the range $q \in [0,10]$, for $R=1$, $\Lambda=5 \times 10^4$ and $M=0$. The red dots come from the numerical integration of \eqref{resto}. The continuous blue curve is $\widetilde R_2(q)$ for the same values of $R$ and $M$ in \eqref{R2}. {\it Right panel}: the same as in the left panel, with $\Lambda=10^2$. With respect to the analogous case of section 3.1, the drift of the red dots from the blue curve (at increasing values of $q$) is more pronunced, and this is why we need a wider range for the vertical axis (note that, despite the difference in scale, the blue curve in the right panel is the same as the blue curve in the left panel).}
	\label{Interpolations}
\end{figure}

To proceed with our analysis, let us consider now the $M=0$ and $q=q(\phi)$ case, i.e.\,the case where $V_{1l}(\phi)$ is given by \eqref{EP2}. Assigning specific values to $R$ and $\Lambda$, from \eqref{resto} and \eqref{B} we can evaluate numerically  the rest $R_2$ as a function of $q$. In the left panel of Fig.\,\ref{Interpolations} the result for $R=1$ and $\Lambda=5\times 10^4$ in the range $q \in [0,10]$ is given by the red dots. The continuous blue curve is the plot of \eqref{R2} for the same value of $R$. 
As it was the case for the cylindrical regularization (section 4.1), under the condition $q^2/\Lambda^2 \ll 1$ also
for the spherical regularization the rest $R_2$  coincides with $\widetilde{R}_2$ in \eqref{R2}  (here we have $M=0$, i.e.\,\,$x=0$), i.e.\,with the usual result \eqref{Va} for $V_{1l}(\phi)$. 

As before, we now investigate on the importance of the condition $q^2(\phi)/\Lambda^2 \ll 1$. Again we consider \eqref{resto} for $R_2(q)$ in the range $q \in [0,10]$ with $R=1$ (as in the left panel of Fig.\,\ref{Interpolations}), but this time with a lower cutoff, $\Lambda=10^2$. The resulting curve is given by the red dots in the right panel of Fig.\,\ref{Interpolations}, while the blue continuous curve (as for the left panel) is the plot of \eqref{R2} for the same value of $R$. We immediately see that with this choice of $\Lambda$ the superposition between the dotted curve and the continuous one holds for a more limited range of $q$. In this case, in fact, the condition $q^2/\Lambda^2 \ll 1$ breaks down for smaller values of $q$.

In conclusion, as long as we limit ourselves to consider values of $q$ much smaller than $\Lambda$ (as we have to), 
the rest $R_2$ coincides with the usual finite result $\widetilde R_2$ in \eqref{R2}. Under this condition, from Eq.\,\eqref{EP2} we obtain 
\begin{align}\label{divpotiq}
	&V_{1l}(\phi)=\frac{3}{180 \pi ^2}\,q^2(\phi)\,R\Lambda^3 -\frac{3}{840 \pi ^2}\,q^4(\phi)\, R \Lambda + \widetilde R_2+\mathcal{O}(\Lambda^{-1})\,,
\end{align}
from which we see that, in addition to the UV-insensitive result $\widetilde{R}_2$ obtained in the usual approach (section 2), UV-sensitive field-dependent terms also contribute to $V_{1l}(\phi)$. 

\begin{figure}[t]
	\centering
	\includegraphics[scale=0.2]{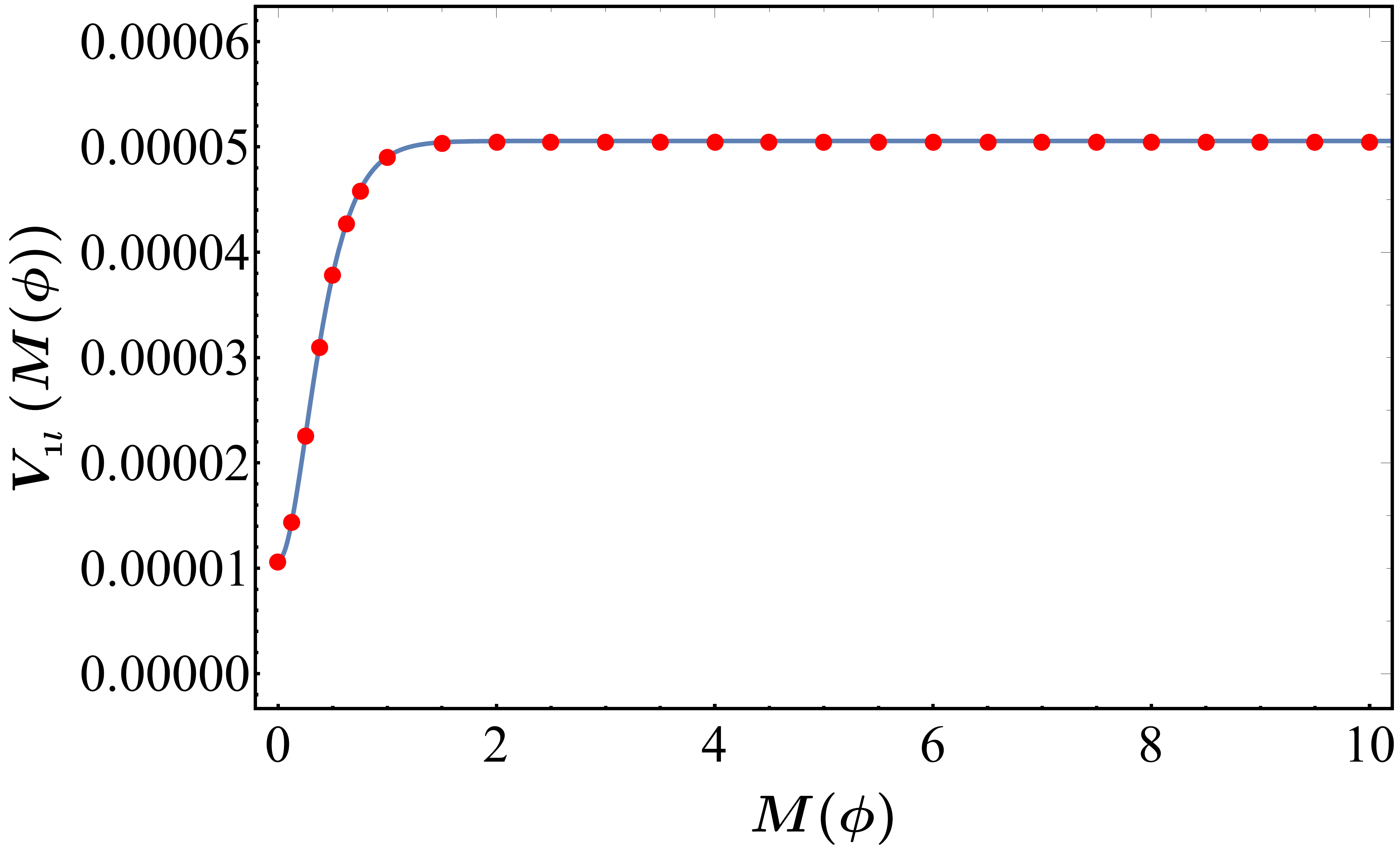}
	\hspace{3mm}
	\includegraphics[scale=0.2]{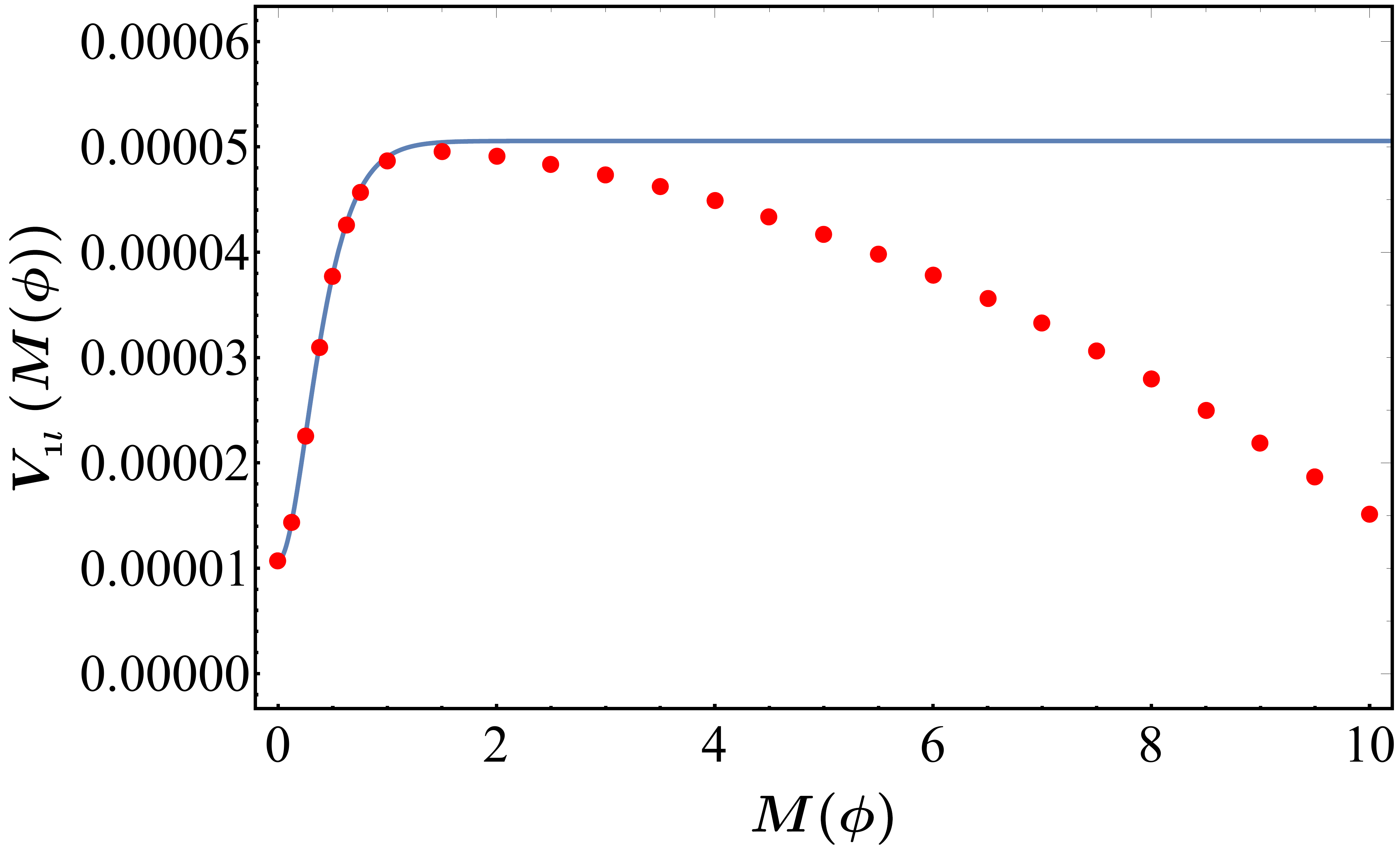}
	\caption{{\it Left panel}: the rest $R_{2r}$ for $r=1$ as a function of $M$ in the range $M \in [0,10]$, for $R=1$, $\Lambda=5\times10^4$ and $q=0.1$. The red dots come from the numerical integration of \eqref{resto}. The continuous blue curve is $\widetilde R_2(M)$ in \eqref{R2} for the same values of $R$ and $q$. {\it Right panel}: the same as in the left panel, with $\Lambda=10^2$.}
	\label{Interpolationms}
\end{figure}

Let us consider now the cases when $V_{1l}(\phi)$ is given by \eqref{EP1}, i.e.\,models with $M^2=M^2(\phi)$ and $q$ field-independent.
Similarly to the previous case, we assign specific values to $R$, $q$ and $\Lambda$, and evaluate numerically the rest $R_2$ in \eqref{resto} as a function of $M$. In the left panel of Fig.\,\ref{Interpolationms} the numerical result for  $R_2(M)$ with $R=1$, $q=0.1$ and $\Lambda=5\times10^4$ in the range $M \in [0,10]$ is given by the red dots, while the continuous blue line is the plot of the analytic curve  $\widetilde{R}_2(M)$ in \eqref{R2} for the same values of $R$ and $q$. 
Under the condition $M^2/\Lambda^2 \ll 1$, once again we see that the rest $R_2$ 
in \eqref{resto} coincides with $\widetilde{R}_2$ in \eqref{R2}: the finite contribution \eqref{R2} to $V_{1l}(\phi)$ is present also with the spherical regularization.

In the right panel of Fig.\,\ref{Interpolationms} we plot again the numerical result for $R_2(M)$ in the range $M \in [0,10]$, where we take $R=1$ and $q=0.1$ (as for the left panel), but $\Lambda=10^2$. The result is given by the red dots, while the blue continuous curve is the plot of \eqref{R2} for the same values of $R$ and $q$. Differently from the left panel, the superposition between dotted  and  continuous curve holds for a more limited range of $M$, as in this case the condition $M^2/\Lambda^2 \ll 1$ breaks down for smaller values of $M$.

Focusing as before on SUSY models, we now consider the contribution to $V_{1l}(\phi)$ from a single couple of superpartners. From \eqref{divpoti} we get 
\begin{align}\label{divpotisusy}
	&V_{1l}(\phi)= -\frac{14 (q_b^2-q_f^2)}{840 \pi ^2}\,M^2(\phi)\, R \Lambda+ \widetilde R_2^{(b)}- \widetilde R_2^{(f)}+\mathcal{O}(\Lambda^{-1}),
\end{align}
from which we see that $V_{1l}(\phi)$ is given by the sum of the usual finite result (section 2), that here consists in the difference $\widetilde R_2^{(b)}- \widetilde R_2^{(f)}$, with a UV-sensitive term. An important outcome of the present analysis is that, while in \eqref{divpoti} the UV-sensitive terms proportional to powers of $M^2(\phi)$ (but independent of $q_i$) are cancelled by the presence of superpartners, the same does not hold for the UV-sensitive terms of the kind $q_i^2 M^2(\phi) R \Lambda$, with the exception of the $q_b=q_f$ irrelevant case.

Let us discuss now the findings of the present section. The result in \eqref{divpoti} (see also \eqref{divpotiq} and \eqref{divpotisusy}) for $V_{1l}(\phi)$ can hardly be overestimated. 
It contains two different contributions: (i) the result obtained in the usual approach (see the last line of \eqref{bosonfinalp} in section 2); (ii)
additional UV-sensitive terms proportional to powers of $q$. These latter terms do not vanish due to the fact that in the calculation of $V_{1l}(\phi)$ we included the asymptotics of the loop momenta in a proper manner. Their absence in the celebrated UV-insensitive result for $V_{1l}(\phi)$ comes from an inconsistent limiting procedure. We have also found that the finite contribution to the potential does not arise because of the infinite sum over $n$, but rather from the fact that we take into account the necessary physical limitations of the calculation, namely $q^2/\Lambda^2,M^2/\Lambda^2 \ll 1$.

It is of the greatest importance to stress that the UV-insensitive contribution to $V_{1l}(\phi)$ (usual result) and the UV-sensitive ones found in the present work
have a totally different origin. 
The former (right-hand side of \eqref{R2}) arises because the fifth component of the loop momentum comes with a sum rather than an integral. 
Physically this is due to the large scale hierarchy between the dimension of the $4D$ box ($\sim \mathbb R^4$) and the finite radius $R$ of $S^1$. The use of the EML formula greatly helps in highlighting this point: the finite term is nothing but the rest of the EML. In this respect, we observe that such a term vanishes in the $R\to \infty$ limit (i.e.\,when the difference between sum and integral is practically washed out). The physical origin of the $q$-dependent UV-sensitive terms is different. Their presence is entirely due to the topology of the non-simply connected $\mathbb R^4\times S^1$ manifold, which implies that boundary conditions are needed to define the theory. In particular, non-trivial boundary conditions as those in  \eqref{nontrivialbc} can be realized, independently of the value of $R$. This holds true even when the size of the $4D$ box and $R$ are comparable, that in turn means even when the sum over the fifth component of the momentum can be replaced with an integral (as it is the case for the other four components). Therefore, these previously missed $q$-dependent UV-sensitive terms are a consequence of the non-trivial topology of the spacetime, and are always present when non-trivial boundary conditions are realized. 

One might think that when the size of the compact dimension is sufficiently large ($R \to \infty$) we should recover for $V_{1l}(\phi)$ the result obtained in $\mathbb R^5$. However we have just seen that, whatever the size of the compact dimensions of the  multiply connected manifold, in the presence of global supersymmetry the physics in $\mathbb R^{4+n}$ can sensibly differ from the physics in $\mathbb R^4 \times H$ (with $H=T^n$, $T^n/Z_2$, \dots, where $T^n$ is the $n$-dimensional torus). The difference is due to the UV-sensitive terms of topological origin that we have found. 
Physically this comes from the difference between $q_b$ and $q_f$: the non-trivial boundary conditions introduce a mismatch in the mode expansion of superpartners. This determines a breaking of SUSY, governed by the parameter $(q_b^2-q_f^2)$, in a way similar to the more typical and better known breaking that arises in $\mathbb R^{4+n}$ when the masses of superpartners do not coincide. Moreover, the fact that in \eqref{divpotisusy} the field-dependent $\Lambda^3$ terms are cancelled seems to suggest that the breaking is spontaneous. In fact, we are left with linearly divergent terms only, that are reminiscent of the typical $\log \Lambda$ terms of a softly broken $4D$ theory (the linear divergence is clearly the upgrade of the logarithmic one when we move from four to five dimensions). On the contrary, the result \eqref{divpotiq} for the supersymmetric models\cite{Barbieri:2000vh} described by \eqref{EP2} shows a cubic divergence, so that one could be tempted to conclude that in this case the SUSY breaking is hard. However, we should note that \eqref{EP2} is obtained only after the diagonalization of an infinite KK mass matrix. According to our findings, such a procedure is ill-defined. In fact, treating the descendant theory as a $4D$ theory with an infinite number of KK-fields corresponds to take the asymptotics of the Fourier component of the momentum along the fifth compact dimension independently from the others, and we have shown that this is an inconsistent procedure.

To further deepen our analysis, in the next section we calculate $V_{1l}(\phi)$ introducing a smooth cutoff.

\section{UV-sensitivity of $\boldsymbol{V_{1l}(\phi)}$. Smooth cutoff}
In the present section we consider the calculation of $V_{1l}(\phi)$
resorting to a smooth regularization. In section 5.1 we introduce a smooth cutoff only for the four components $p=(p_1,p_2,p_3,p_4)$ of the $5D$ momentum. The reason for this choice is that we want to check whether, treating again the sum over $n$ independently from the integration over $p$, even with a smooth cutoff we recover the finite result of the usual calculation. The consistent way of implementing the asymptotics of the $5D$ momentum $p^{(5)}$ is considered in section 5.2, where a smooth spherical cutoff for the $5D$ loop momenta $p^{(5)}=(p,n/R)$ is introduced.

\subsection{Smooth cutoff on $\boldsymbol{p=(p_1,p_2,p_3,p_4)}$}
We now calculate $V_{1l}(\phi)$ introducing a smooth cutoff for the integration over $p$ through the cutoff function $e^{-p^2/\Lambda^2}$, and (as in section 2) we perform first the infinite sum over $n$ and then the integration over $p$.
 
Referring to Eq.\,\eqref{ellipticthetaintp1} for a bosonic contribution to  $V_{1l}(\phi)$,  multiplying the integrand in \eqref{ellipticthetaintp1} for $e^{-p^2/\Lambda^2}$, and using the Poisson summation formula we get
\begin{align}\label{Vinfinitesumsmooth}
	V^b_{1l}(\phi)&=-\frac{1}{2} \int \frac{d^4p}{(2\pi)^4}e^{-\frac{p^2}{\Lambda^2}}\int_0^\infty \frac{ds}{s}\sum_{n=-\infty}^{\infty} \left[e^{-s(R^2(p^2+M^2)+(n+q_b)^2)}-e^{-s\,(p^2R^2+n^2)}\right] \nonumber \\
	&=-\frac{1}{2} \int  \frac{d^4p}{(2\pi)^4} e^{-\frac{p^2}{\Lambda^2}}\int_0^\infty \frac{ds}{s^{3/2}} \sqrt{\pi } \left[ \vartheta _3\left(\pi  q_b,e^{-\frac{\pi ^2}{s}}\right) e^{-s R^2 \left(p^2+M^2\right)}-  \vartheta _3\left(0,e^{-\frac{\pi ^2}{s}}\right)e^{-s R^2p^2}\right], 
\end{align}
where $\vartheta_3(x,y)=1+2 \sum_{k=1}^{\infty} \cos(2k x) y^{k^2}$ was already given in \eqref{theta3}.

As in section 2, we begin by taking in \eqref{Vinfinitesumsmooth} only the term \vv $1$" for both of the $\vartheta_3$, thus obtaining \eqref{ints}. Performing successively the integration over $p$ we get
\begin{align}\label{divm2}
R \left(\frac{\Lambda ^3 M^2}{64 \sqrt\pi }-\frac{\Lambda M^4}{128 \sqrt\pi }+\frac{M^5}{60 \pi }\right) +O\left(\frac{1}{\Lambda}\right),
\end{align}
where we expanded the result for $M/\Lambda\ll 1$. Comparing  \eqref{divm2} with its hard cutoff counterpart \eqref{divm}, we see that the only difference between these two results is in the value of the numerical coefficients in front of the divergent terms\footnote{It is worth to stress that the coefficients of (non-universal) power-like divergences obtained through different regularization procedures may well differ.
Moreover, we know that in four-dimensional QFTs the coefficients of the logarithmic divergences are universal and do not depend on the regularization. In fact they determine the anomalous dimension of the couplings, that is in turn related to the critical exponents. In the present $5D$ case the term that shows universality is the finite $M^5$ term. It has the same coefficient in \eqref{divm}, \eqref{divpot}, \eqref{divpoti} and \eqref{divm2}. The reason is that the $M^5$ contribution is generated by the sum over $n$ of terms of the kind $(m_n^2)^2\log(m_n^2)$ (with $m_n^2\equiv M^2+(n/R+q)^2$), that are well known to be universal.}.

Let us go on with our calculation, and consider the first of the two $\vartheta_3$ in \eqref{Vinfinitesumsmooth}. Inserting in this equation the remaining part of the $\vartheta_3$ function (the series of cosines), for this contribution to $V_{1l}(\phi)$ we get
\begin{align}
	\label{KK result smooth}
	&\sum_{k=1}^{\infty}\Big[-\frac{e^{-2 \pi k M R} (2 \pi k M R (2 \pi k M R+3)+3) \cos (2 \pi  k q_B)}{64 \pi ^6 k^5 R^4} \nonumber\\
	&-\frac{e^{-2 \pi  k M R} \left(-8 \pi ^3 k^3 M^3 R^3-24 \pi ^2 k^2 M^2 R^2-30 \pi  k M R-15\right) \cos (2 \pi  k q)}{64 \Lambda ^2 \left(\pi ^8 k^7 R^6\right)}+\dots\Big]\,,
\end{align}
where the second term in the square brakets and the terms indicated with the dots are suppressed in the $\Lambda \to \infty$ limit. Comparing with the corresponding hard cutoff result \eqref{litium}, we see that the finite results in \eqref{KK result smooth} and \eqref{litium} coincide\footnote{All the other terms (that of course vanish in the $\Lambda \to \infty$ limit) are suppressed as inverse powers of $\Lambda$, while in the hard cutoff case \eqref{litium} they are exponentially suppressed by factors of the kind $e^{-2\pi k R \Lambda}$.}.

To complete our calculation, we still have to consider the remaining term in the second line of \eqref{Vinfinitesumsmooth}, namely the contribution of the cosines in the second of the two $\vartheta_3$ functions. We get
\begin{align}\label{finiteterm smooth}
	\sum_{k=1}^\infty\int \frac{d^4p}{(2\pi)^4} e^{-\frac{p^2}{\Lambda^2}}\int_0^{\infty}\frac{ds}{s^{3/2}}\sqrt{\pi } e^{-s R^2 p^2} e^{-\frac{\pi ^2 k^2}{s}} =\sum_{k=1}^\infty \frac{3}{64 \pi ^6 k^5 R^4}+\mathcal{O}\left(\frac{1}{\Lambda^2}\right)\simeq\frac{3 \zeta (5)}{64 \pi ^6 R^4},
\end{align}
that coincides with the hard cutoff result \eqref{finterm}.

Putting together \eqref{divm2}, \eqref{KK result smooth} and \eqref{finiteterm smooth}, we get the complete result for $V_{1l}(\phi)$. Once again we see that, having introduced a cut (although smooth) over $p$ but none for $n$,  the usual finite result (section 2) for $V_{1l}(\phi)$ is obtained. The reason
is that even in this case we treated the asymptotics of the different components of $p^{(5)}$ in an asymmetric (incorrect) manner.

\subsection{Smooth cutoff over the $\boldsymbol 5$-momentum $\boldsymbol{p^{(5)}=(p,n/R)}$}
In this section we calculate $V_{1l}(\phi)$ using the $5D$ cutoff function $e^{-\frac{p^2+n^2/R^2}{\Lambda^2}}$, that smoothly suppresses in the loop the contribution of the modes with $5D$ momenta $(p^{(5)})^2= p^2+n^2/R^2 \geq \Lambda^2$:
\begin{equation}\label{pot smooth cutoff}
	V_{1l}(\phi)=\sum_{n=-\infty}^\infty\int \frac{d^4p}{(2\pi)^4}\left\{\log\left(p^2+M^2+\left(\frac{n}{R}+q\right)^2\right)-\log\left(p^2+\frac{n^2}{R^2}\right)\right\}e^{-\frac{p^2+n^2/R^2}{\Lambda^2}}.
\end{equation}

In connection with the warnings raised in previous sections (correct treatment of the $p^{(5)}$ asymptotics), it is important to observe that the presence in \eqref{pot smooth cutoff} of the cutoff function $e^{-(p^2+n^2/R^2)/\Lambda^2}$ makes the integrand sufficiently damped, so that the infinite sum over $n$ and the integration in the infinite domain $\mathbb R^4$ over $p$ can be performed in any order, obtaining always the correct result. From the mathematical point of view, this means that in this case there is no need to pay attention to the construction of the set of compact domains.

Comparing with the hard cutoff calculation of section 4.2, the upside of the present regularization is that it manifestly respects the higher-dimensional symmetries of the model, as for e.g.\,the $5D$ Lorentz symmetry (before euclideanization). Moreover, thanks to the fact that we can perform the infinite sum over $n$ without making any violence to the asymptotics of $p^{(5)}$, the above expression also preserves the shift symmetry along the circle. In fact, while implementing a decoupling of the modes with $(p^{(5)})^2\ge \Lambda^2$, the potential $V_{1l}(\phi)$ in \eqref{pot smooth cutoff} is invariant under the transformation $n\to n+m$ (with $m$ integer).

Performing in \eqref{pot smooth cutoff} the integration over $p$, we get 
\begin{align}\label{Gn}
	V_{1l}(\phi)=&\sum_{n=-\infty}^{\infty}\frac{\Lambda ^2e^{-\frac{n^2/R^2}{\Lambda ^2}}}{32 \pi ^2} \Bigg[e^{\frac{\left({n}/{R}+q\right)^2+M^2}{\Lambda ^2}} \left(M^2+\left(\frac{n}{R}+q\right)^2-\Lambda ^2\right) \text{Ei}\left(-\frac{\left(\frac{n}{R}+q\right)^2+M^2}{\Lambda ^2}\right) \nonumber\\
	&+\Lambda ^2 \log \left(\frac{\left(\frac{n}{R}+q\right)^2+M^2}{n^2/R^2}\right)-e^{\frac{n^2/R^2}{\Lambda ^2}} \left(\frac{n^2}{R^2}-\Lambda ^2\right) \text{Ei}\left(-\frac{{n^2}/{R^2}}{\Lambda ^2}\right)\Bigg] 
	\equiv \sum_{n=-\infty}^{\infty}G(n),
\end{align}
where ${\rm Ei} (x)$ is the exponential integral function. Resorting again to the EML formula, we have
\begin{align}\label{EMLsmooth}
	V_{1l}(\phi)=\lim_{y\to\infty}\left[\int_{-y}^y dx\, G(x)+\frac{G(y)+G(-y)}{2}+\sum_{k=1}^{r}\frac{B_{2k}}{(2k)!}\left(G^{(2k-1)}(y)-G^{(2k-1)}(-y)\right)+R_{2r}\right],
\end{align} 
where the rest $R_{2r}$ is given by (see \eqref{resto})
\begin{equation}\label{restosmooth}
	R_{2r}=\frac{(-1)^{2r+1}}{(2r)!}\int_{-y}^{y}dx\,G^{(2r)}(x)B_{2r}(x-[x]),
\end{equation}
with $B_{2k}$ and $B_{2r}(x-[x])$ defined below equations \eqref{EML} and \eqref{resto}.

As $G(y)$ and its derivatives all vanish in the $y\to\infty$ limit, in \eqref{EMLsmooth} the terms beside the integral and the rest  all give null  contributions to $V_{1l}(\phi)$. This results from the presence of the damping factor $e^{-n^2/(R^2\Lambda^2)}$ in \eqref{Gn}. 
Two different choices of $r$ for the calculation of the rest differ only for terms that vanish in the $y\to\infty$ limit, so that, depending on calculation convenience, one can choose which value of $r$ to consider.
Unfortunately, the integral in \eqref{EMLsmooth} cannot be performed analytically. This might seem a serious obstacle to our calculation, but we will see that it is possible to overcome such an apparent problem: combining together analytic and numerical calculations, we will eventually obtain the result.  

Let us expand first $G(x)$ in \eqref{EMLsmooth} in powers of $M$ and $q$, and then evaluate the integral. We obtain
\begin{equation}
	\label{V smooth expansions}
	V_{1l}(\phi)=\frac{5 M^2+ 3q^2}{240 \pi ^{3/2}}R\Lambda ^3-\frac{35 M^4+14 q^2 M^2+3 q^4}{1680 \pi ^{3/2}}R\Lambda  +R_2 +\mathcal O\left(\Lambda^{-1}\right)\,.
\end{equation}
From the numerical evaluation of \eqref{restosmooth}, we see that the rest $R_2$ is nothing but $\widetilde R_2$ in \eqref{R2} (see also the left panels of Figs.\,\ref{Interpolations} and \ref{Interpolationms}). Moreover, 
comparing \eqref{V smooth expansions} with its spherical hard cutoff analog \eqref{divpoti}, we see that they have the same kind of divergences with the same relative weight among the terms that multiply a given divergence. Considering for instance the $\Lambda^3$ divergence, we see that \eqref{V smooth expansions} differs from \eqref{divpoti} only for an overall numerical factor. The same holds true for the $\Lambda$ divergence.  

\begin{figure}
\centering	\includegraphics[scale=0.5]{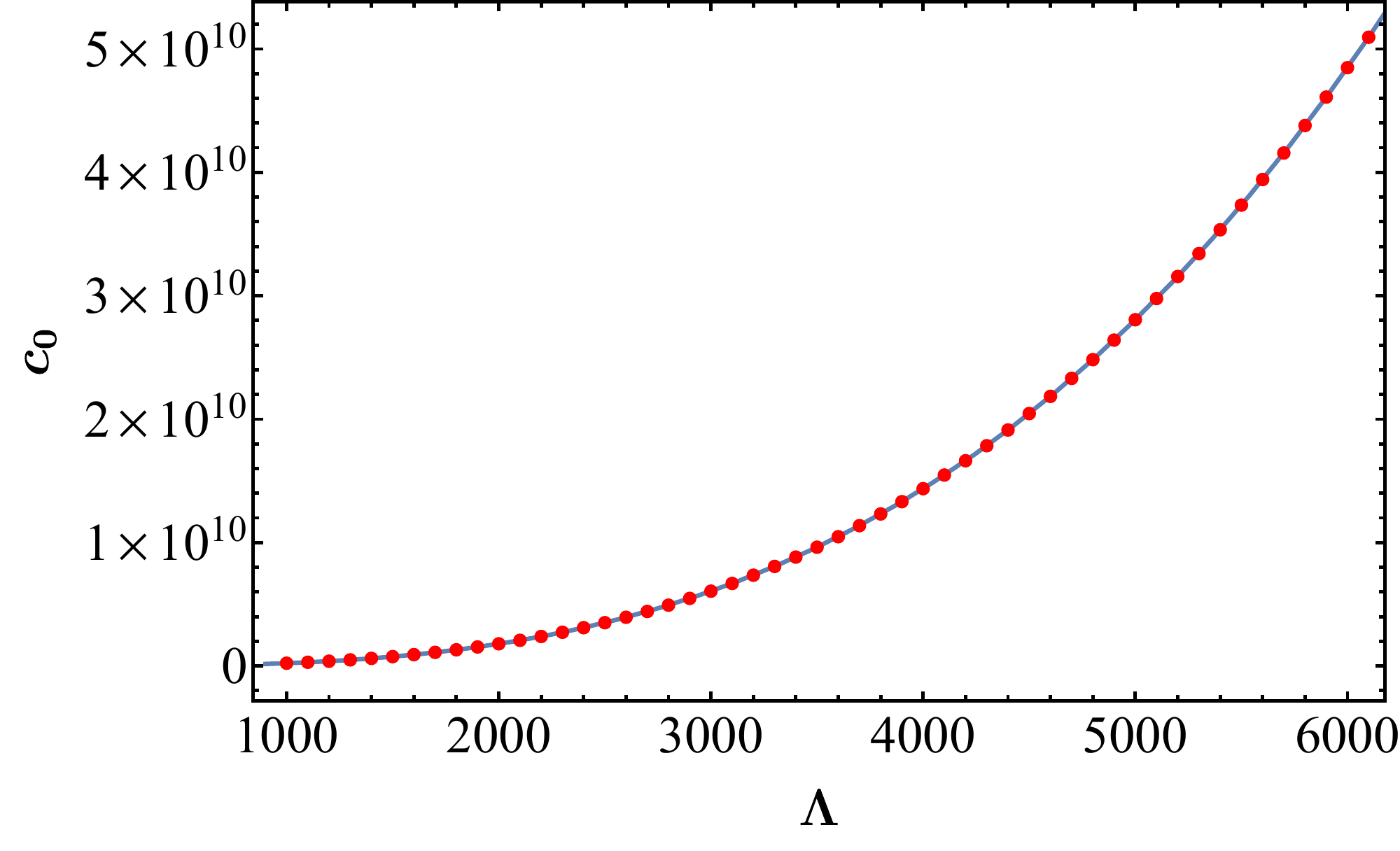}
	\caption{ 
		The red points show the numerical value of the coefficient $c_0$ in \eqref{Mexp} for $R=1$ and $q=10$ as a function of the smooth cutoff $\Lambda$. The blue continuous line is the theoretical prediction given in \eqref{V smooth expansions}, namely $c_0(\Lambda)={q^2 R \Lambda^3}/{80 \pi ^{3/2}}-{q^4 R \Lambda}/{560 \pi ^{3/2}}$. 
	}
	\label{mpower0}
\end{figure}

There is however an important difference between \eqref{divpoti} and \eqref{V smooth expansions}, namely the finite term $M^5$, which is missing in the latter. We then ask ourselves the following question. Is this term missing as it does not appear when a smooth cutoff is used, or is it lost due to the fact that we possibly performed an illegitimate exchange in \eqref{EMLsmooth} between the integral over $x$ and the expansion in powers of $M$ and $q$ in  the term $\int_{-\infty}^\infty dx\, G(x)$\,?
To answer this question, we begin by observing that in this integral, upon expansion of $G(x)$ in powers of $M$, only even powers of $M$ can be generated (the same is true for $q$). Therefore, performing the integral after this expansion, no odd powers of $M$ (and $q$) can ever show up in the final result. This already suggests that making the expansion in $G(x)$ before performing the integral over $x$ might be a mathematically illegitimate operation.

\begin{figure}[t]
	\includegraphics[scale=0.429]{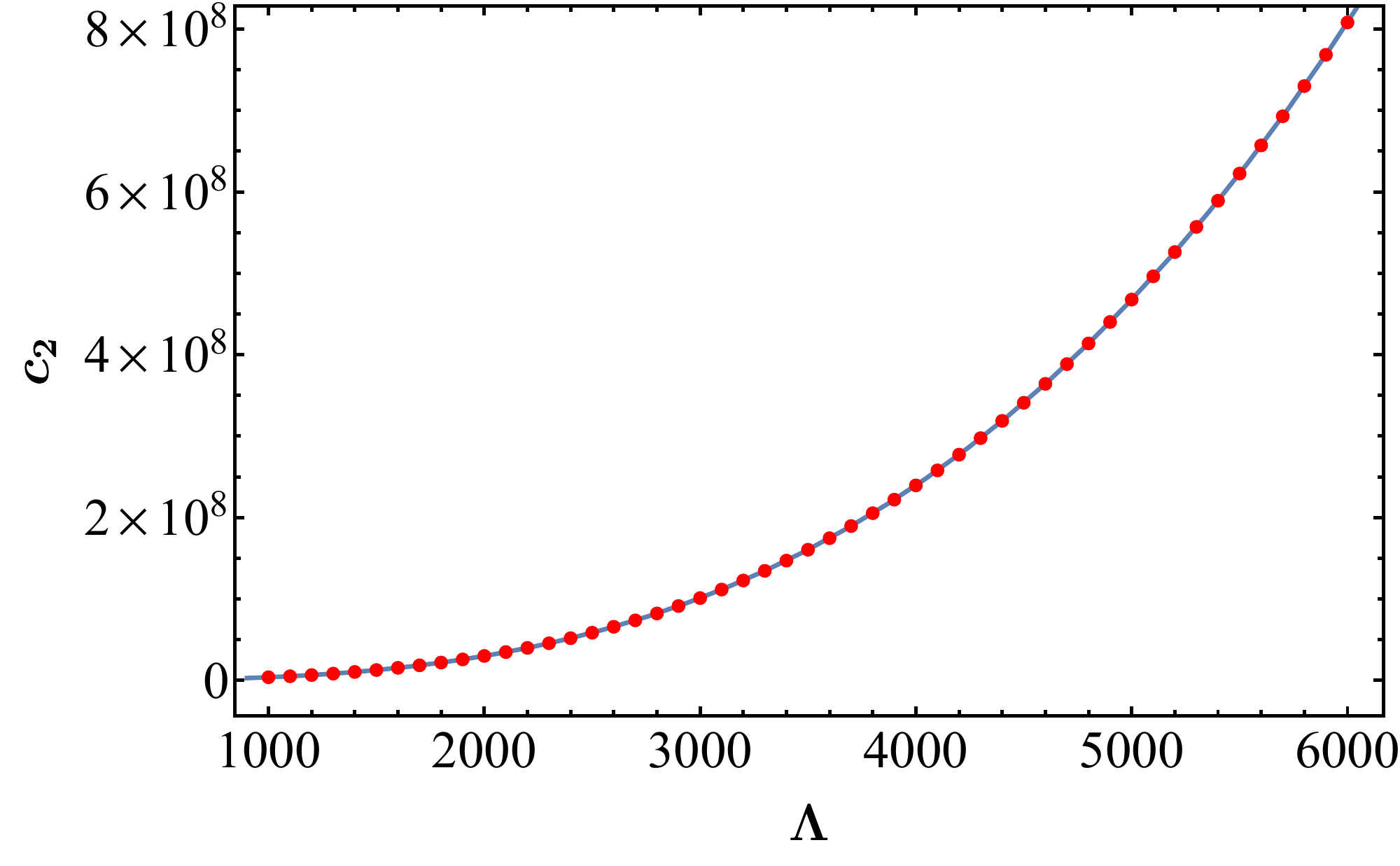}
	\includegraphics[scale=0.471]{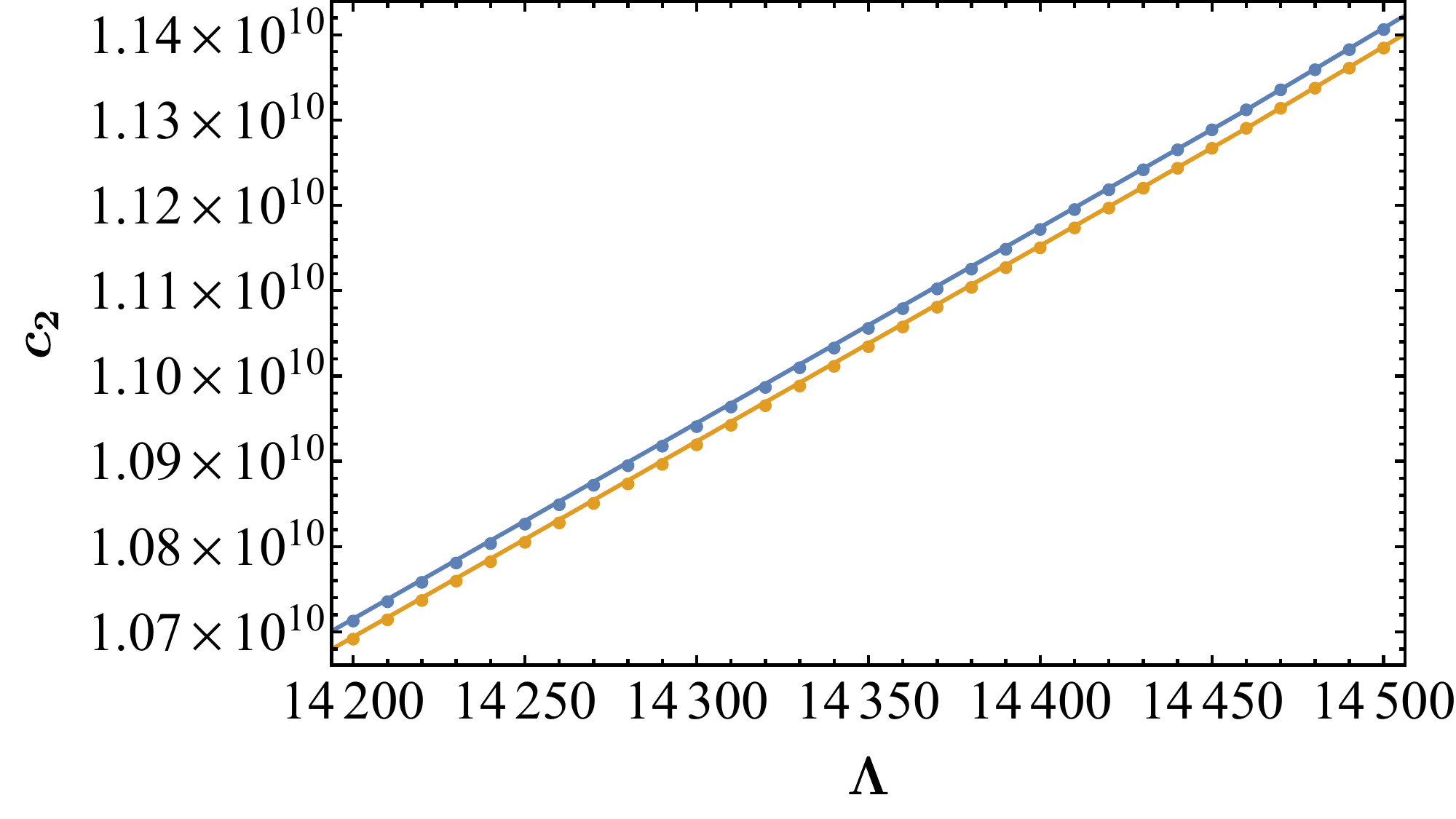}
	\caption{\textit{Left panel}: 
		The red dots show the numerical value of the coefficient $c_2$ in \eqref{Mexp} for $R=1$ and $q=10$ as a function of the smooth cutoff $\Lambda$. The blue continuous line is the theoretical prediction given in \eqref{V smooth expansions}: $c_2(\Lambda)={R\Lambda^3}/{48 \pi ^{3/2}}-{q^2 R\Lambda}/{120 \pi ^{3/2}}$. 
		\textit{Right panel}: Zoom of $c_2$ in the range $14200 \leq \Lambda \leq 14500$ for $q=0.1$ (the blue points show the numerical value, the blue line the analytic curve) and $q=1000$ (yellow points and yellow line).}
	\label{mpower2}
\end{figure}

\begin{figure}[t]
	\includegraphics[scale=0.44]{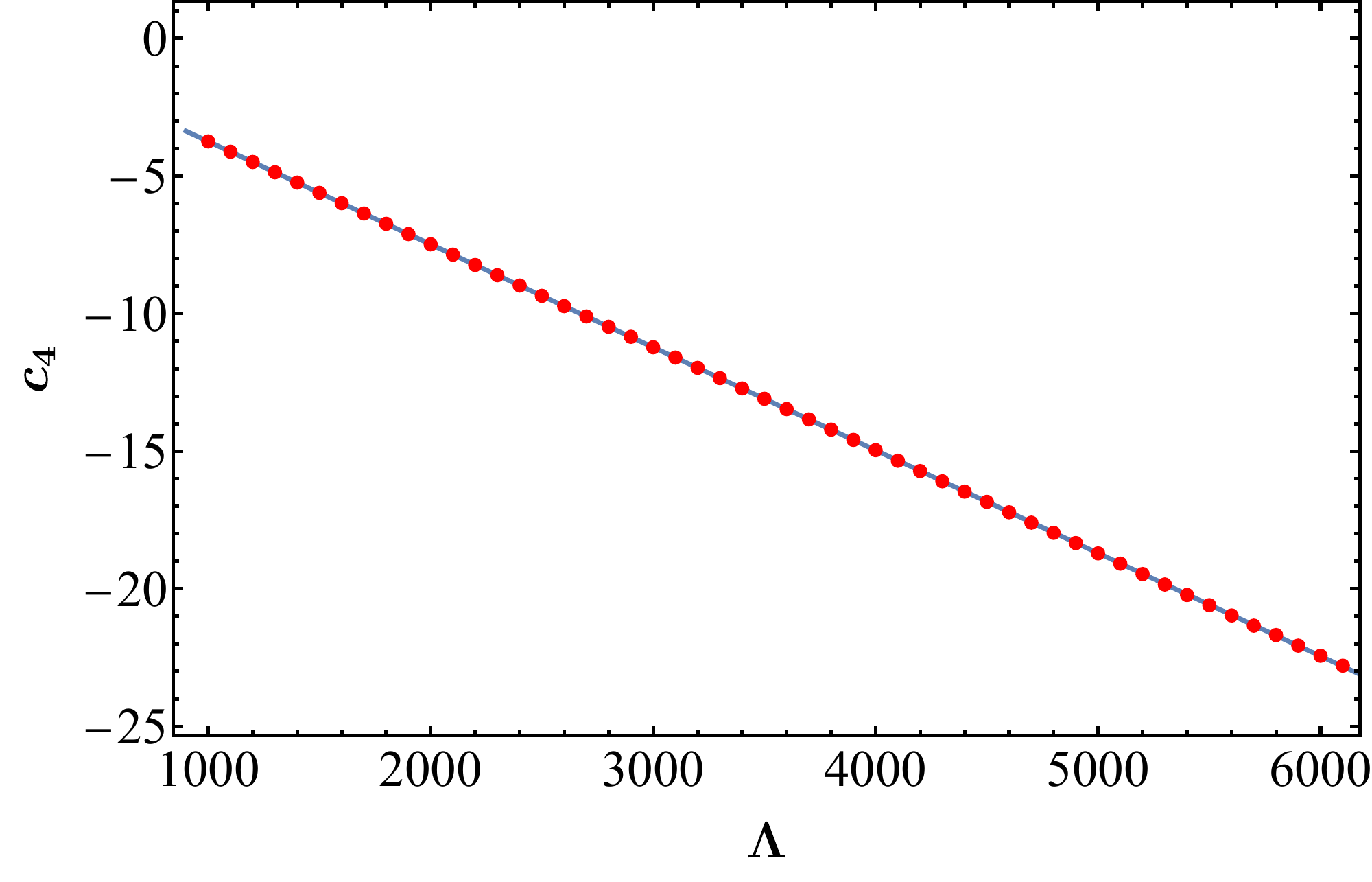}
	\includegraphics[scale=0.44]{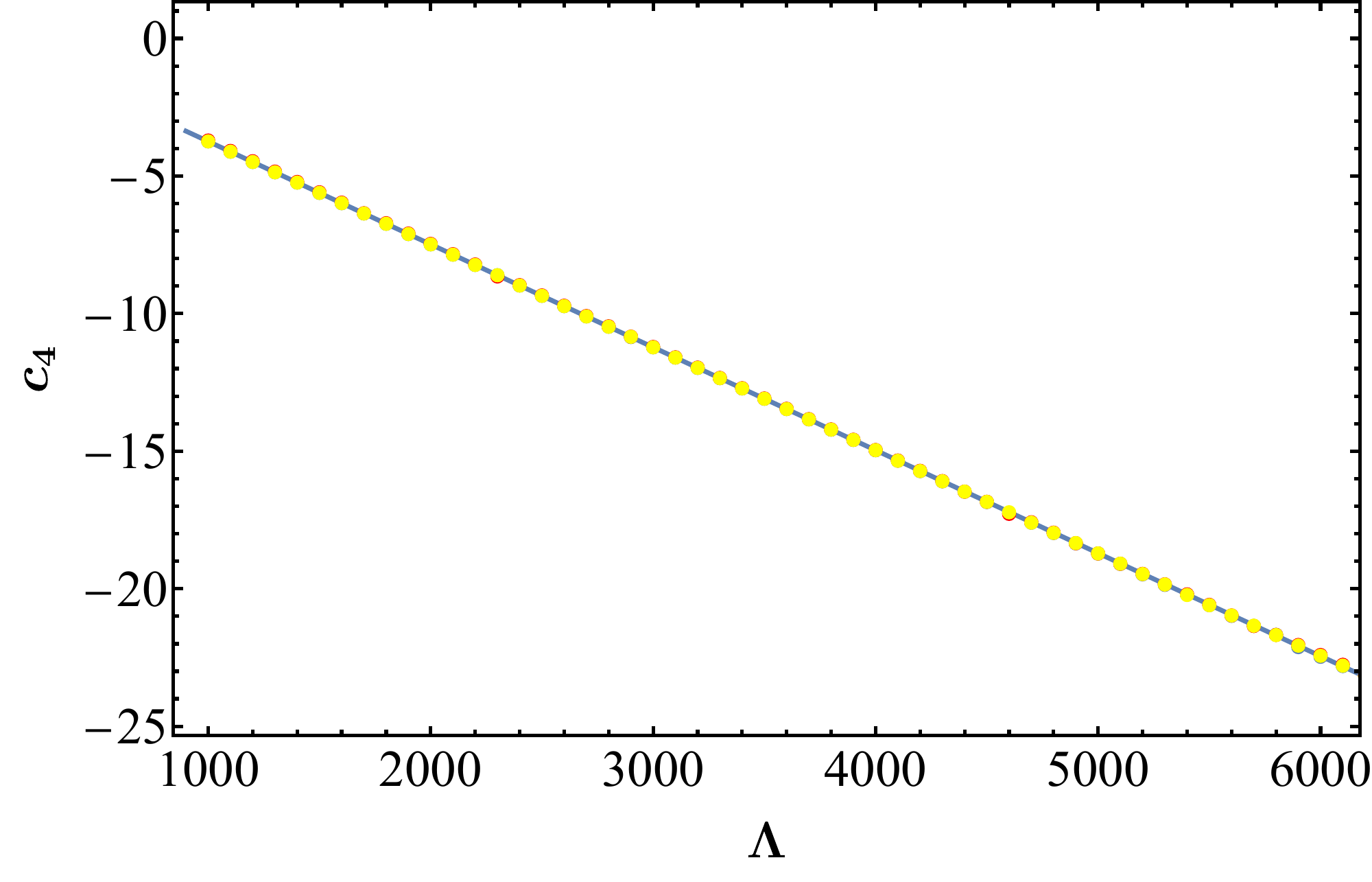}
	\caption{\textit{Left panel}: The red dots show the numerical values of $c_4$ for $q=10$, while the blue continuous line is the theoretical prediction given in \eqref{V smooth expansions}: $c_4(\Lambda)=-{R\Lambda}/{48 \pi ^{3/2}}$. \textit{Right panel}: The red, yellow and green dots are the values of $c_4$ for $q=1$, $q=10$  and $q=100$ respectively. 
		The blue continuous line is again the analytic prediction given above.}
	\label{mpower4}
\end{figure}

\begin{figure}[t]
	\centering
	\includegraphics[scale=0.5]{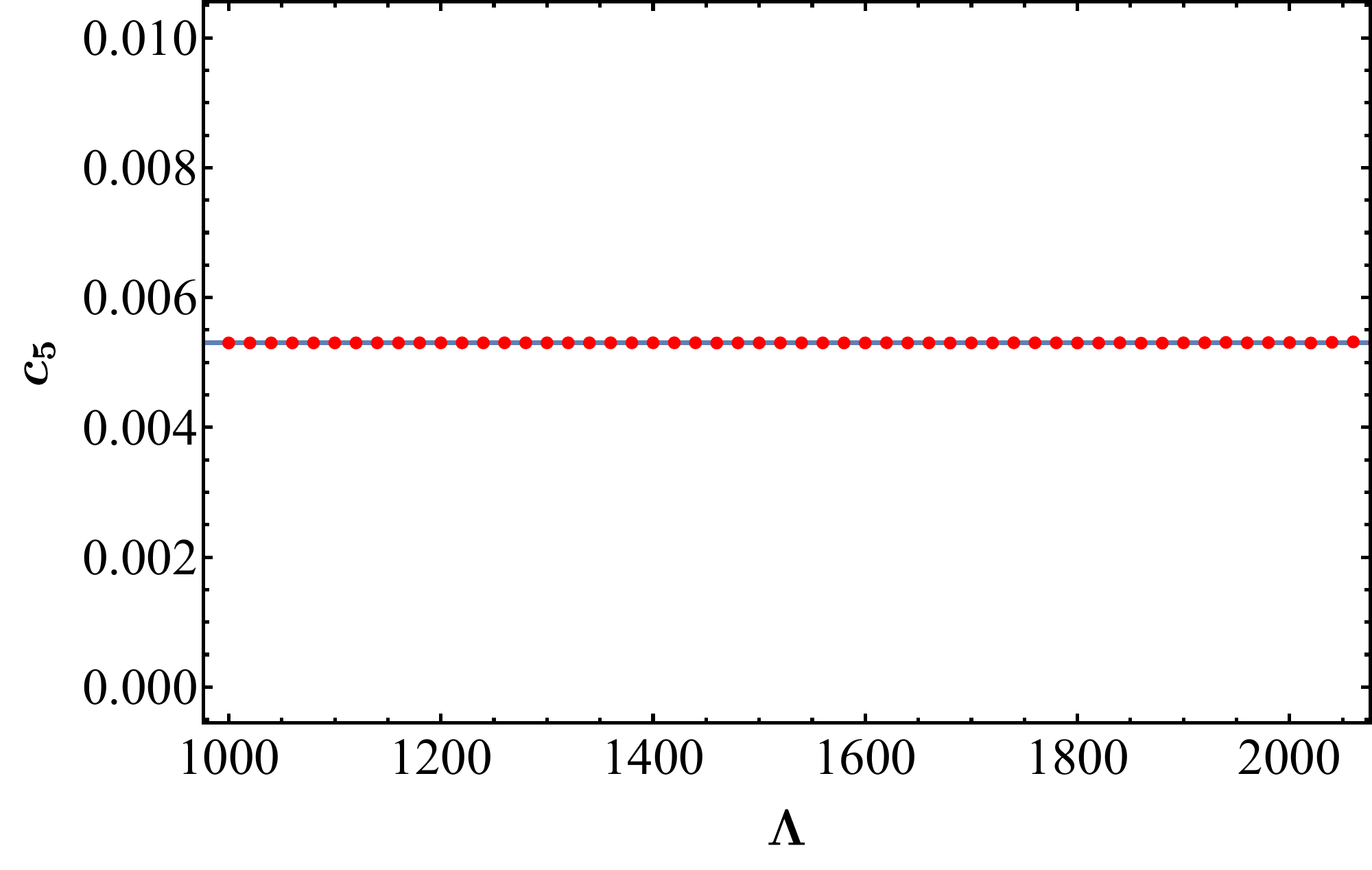}
	\caption{The red dots show the numerical values of $c_5$ as a function of $\Lambda$ for $R=1$ and $q=10$, the blue continuous line is the analytic prediction \eqref{divpoti} of the spherical hard cutoff in section 4.2, $c_5=1/60 \pi $.}
	\label{mpower5}
\end{figure}

To further investigate on this point, we now move to a fully numerical analysis: we evaluate the integral in \eqref{EMLsmooth} numerically, with no reference to any expansion.  
This analysis should allow to check: (i) whether odd powers of $M$ and $q$ are actually present, in particular the term $M^5$; (ii) whether the coefficients of the even powers of $M$ and $q$ in \eqref{V smooth expansions} (obtained after the expansion of $G(x)$) are the correct ones.
To isolate the different powers of $M$ and $q$, we use the following numerical strategy. We first take specific values for $q$ and $R$. Then, given a value of $\Lambda$ (such that the conditions $q,1/R \ll \Lambda$ are fullfilled), we perform the integration over $x$ for a range of values of $M$ such that $M \ll \Lambda$. For any fixed triple $(q,R,\Lambda)$, the integral $I=\int_{-\infty}^\infty dx\, G(x)$ becomes a function of $M$, $I=I(M)$. Successively we fit $I(M)$ with a  polynomial expansion
\begin{equation}\label{Mexp}
I(M)=\sum_{n=0}^6c_n M^n
\end{equation}
and extract the coefficients of each power of $M$. Starting over and over again, while keeping fixed values for $q$ and $R$ but varying $\Lambda$, we obtain the dependence on $\Lambda$ of the various $c_n$: $c_n=c_n(\Lambda)$. 
The results of the fits for the different $c_n(\Lambda)$ are presented in figures \ref{mpower0}-\ref{mpower5}. 

In Fig.\,\ref{mpower0} the coefficient $c_0(\Lambda)$ is plotted in the range $10^3 \leq \Lambda \leq{6 \times 10^3}$, for $R=1$ and $q=10$. The red dots are the outcome of the numerical analysis, while the blue continuous line is the analytic result contained in \eqref{V smooth expansions}, namely  $c_0(\Lambda)=q^2 R \Lambda^3/80 \pi ^{3/2}-q^4 R \Lambda/560 \pi ^{3/2}$. The numerical values perfectly sit on the analytic curve.
The function $c_2(\Lambda)$, again for $R=1$ and $q=10$, is plotted in the left panel of Fig.\,\ref{mpower2}, where again  we see that the numerical values (red dots) sit on the theoretical curve $c_2(\Lambda)={R\Lambda^3}/{48 \pi ^{3/2}}-{q^2 R\Lambda}/{120 \pi ^{3/2}}$ (blue continuous line) contained in \eqref{V smooth expansions}. In order to evidentiate the two different contributions to $c_2(\Lambda)$ (one proportional to $\Lambda^3$ and $q$-independent, the other proportional to $\Lambda$ and $q$-dependent), in the right panel of Fig.\,\ref{mpower2} we compare the two cases $q=0.1$ and $q=10^3$, and zoom in the region $14200 \leq \Lambda \leq 14500$. The distance between the blue dots (curve) and yellow dots (curve) allows to ascertain that the results of the numerical analysis actually contain for $c_2$ the sum of the two analytic terms that appear in the curve $c_2(\Lambda)$ given above. 

In the left panel of Fig.\,\ref{mpower4}, taking once again $R=1$ and $q=10$, we plot the results of the numerical analysis for the function $c_4(\Lambda)$, together with the analytic prediction $c_4(\Lambda)=-{R\Lambda}/{48 \pi ^{3/2}}$. Once more we see that numerical and analytic curves coincide. We also observe that the analytic prediction for $c_4(\Lambda)$, namely $c_4(\Lambda)=-{R\Lambda}/{48 \pi ^{3/2}}$, does not depend on $q$. The numerical investigatation on this point is carried out performing the analysis for $q=1$, $q=10$, and $q=100$. The results are reported in the right panel of Fig.\,\ref{mpower4}. The analysis confirms the independence of $c_4$ from $q$.  

Up to now we considered only even powers of $M$. Moving to the coefficients $c_1$ and $c_3$ in \eqref{Mexp}, and performing for them the same numerical analysis as for $c_0$, $c_2$ and $c_4$, we find that these terms are not present: $c_1=c_3=0$. 
We now come to  investigate on the possible presence of a term proportional to $M^5$. In section 4.2 we have seen that such a term shows up when a {\it hard}  spherical cutoff is used,  but it is missing in \eqref{V smooth expansions}, where not only we implement a {\it smooth} spherical cutoff, but also perform an expansion of $G(x)$ in powers of $M$ and $q$ before proceeding to the evaluation of  $\int_{-\infty}^\infty dx \, G(x)$ in \eqref{EMLsmooth}. The results of our numerical analysis for the function $c_5(\Lambda)$ are reported in Fig.\,\ref{mpower5}. The {red dots} are the numerical values of $c_5$ (obtained for $R=1$ and $q=10$) for different values of $\Lambda$, while the {blue horizontal line} is the analytic prediction obtained with the hard cutoff in \eqref{divpoti}, namely $c_5={1}/{60\pi}$. Repeating the same analysis for different values of $q$ we always obtain the same result for $c_5(\Lambda)$. 

These results allow us to ascertain that: (i) the $M^5$ term is present even when the theory is regularized with a smooth spherical cutoff (we have already seen in \eqref{divpoti} that it is present when an hard spherical cutoff is used); (ii) its absence from \eqref{V smooth expansions} is certainly due to the fact that the expansion of $G(x)$ in powers of $M$ and $q$ before performing the integral over $x$ is illegitimate (in other words, we cannot invert the integral over $x$ with the series in powers of $M$ and $q$, as done to obtain \eqref{V smooth expansions}); (iii) the coefficient of $M^5$ is the same both for the hard and the smooth spherical cutoff regularizations, and does not depend neither on $\Lambda$, nor on $q$. Once again, this indicates the universality of this term.
Finally, pursuing the numerical analysis to consider powers of $M$ higher than $5$, we find that no such terms are present. They all vanish.

\begin{figure}[t]
	\includegraphics[scale=0.425]{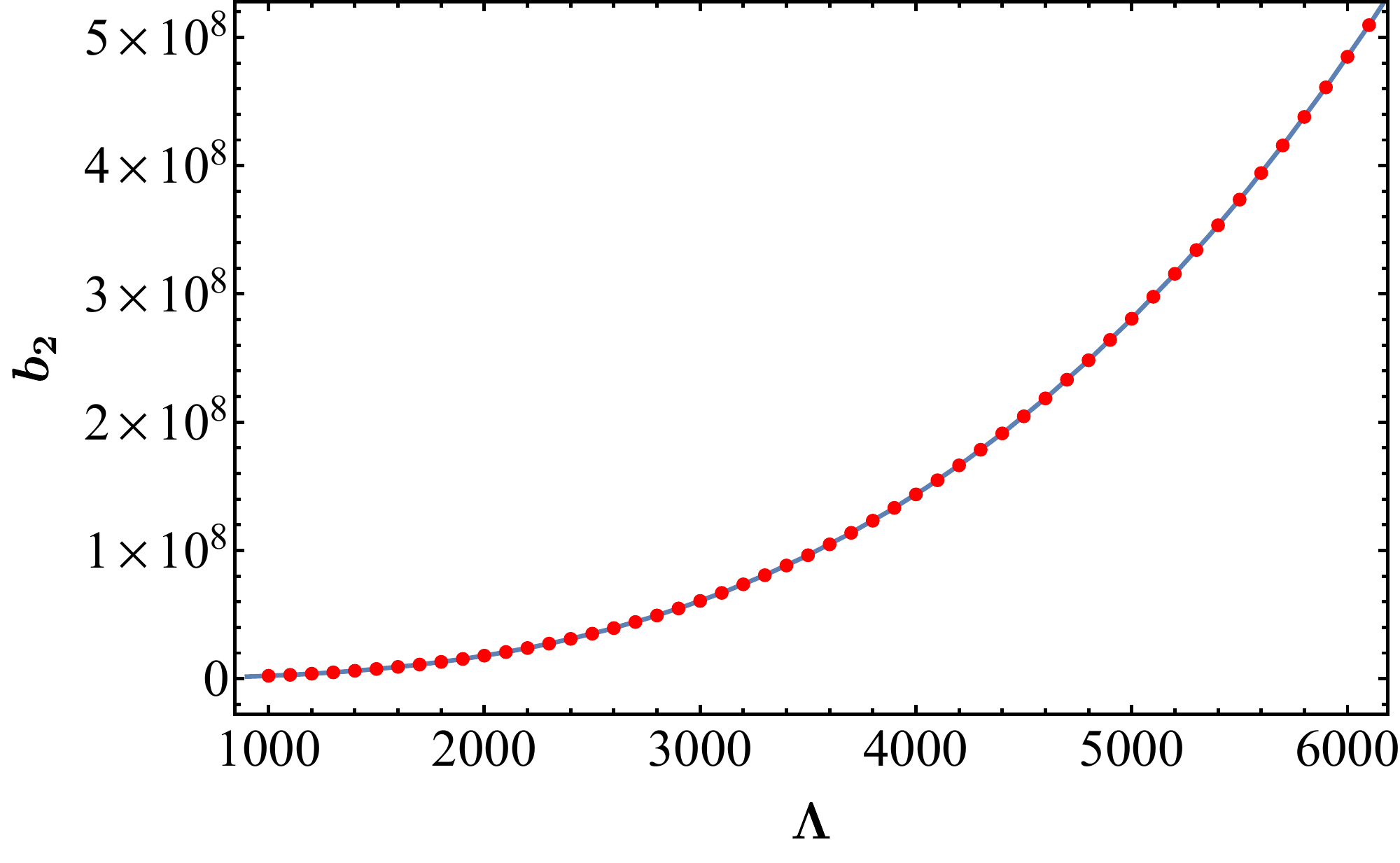}
	\includegraphics[scale=0.455]{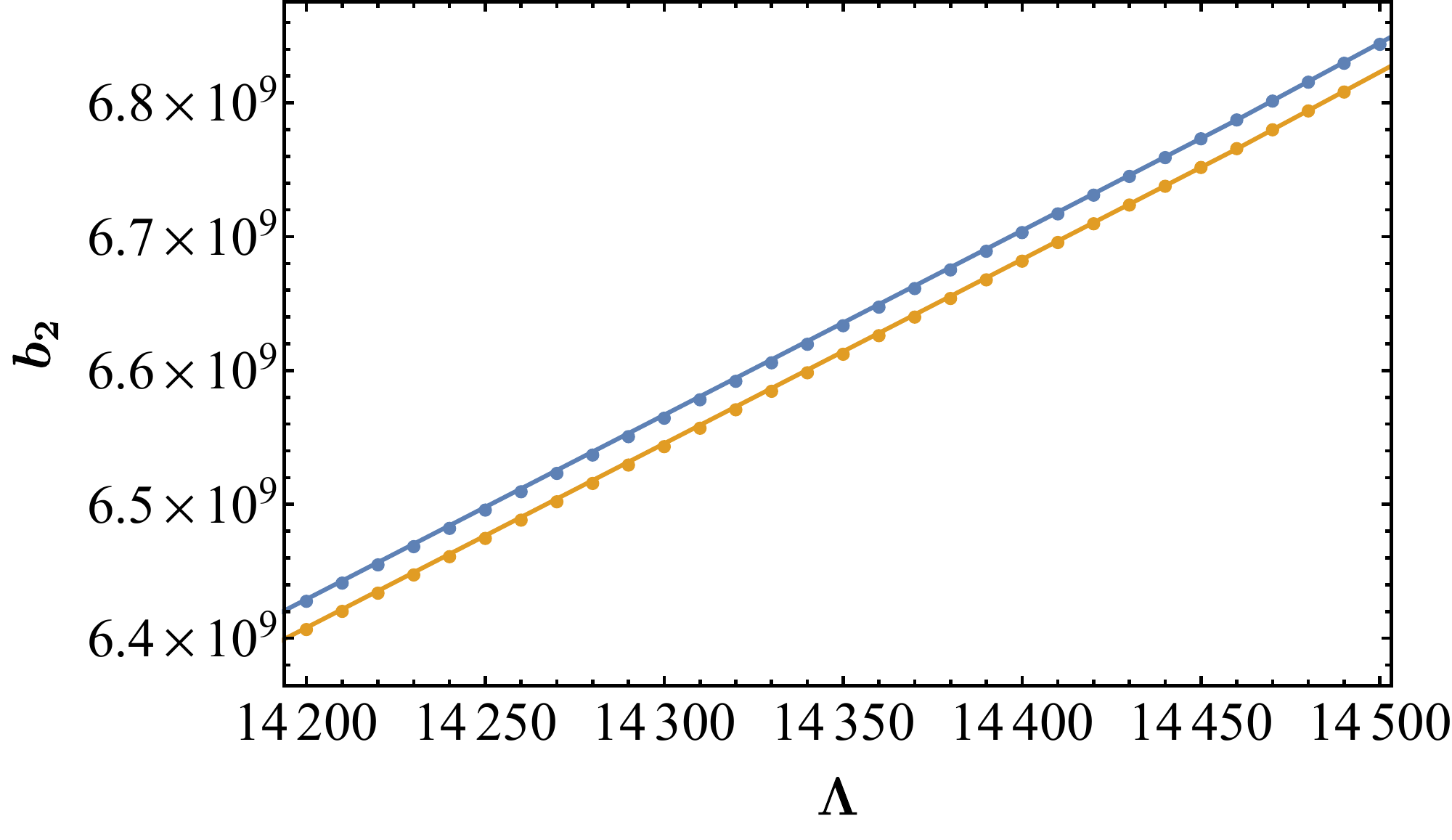}
	\caption{\textit{Left panel}: the red dots are the numerical values of the coefficient of $q^2$, $b_2(\Lambda)$, for $R=1$ and $M=10$. The values of $\Lambda$ are in the range $[10^3,6\times 10^3]$; the blue continuous line is the analytic prediction $b_2(\Lambda)={\Lambda^3}/{80 \pi ^{3/2}}-{M^2 \Lambda}/{120 \pi ^{3/2}}$ contained in \eqref{V smooth expansions}. \textit{Right panel}: Zoom in a limited range of $\Lambda$ of $ b_2(\Lambda)$, for $M=0.1$ (the blue dots show the numerical values, the blue line is the analytic curve) and $M=1000$ (yellow dots and yellow line).}
	\label{qpower2}
\end{figure}

\begin{figure}[t]
	\centering
	\includegraphics[scale=0.5]{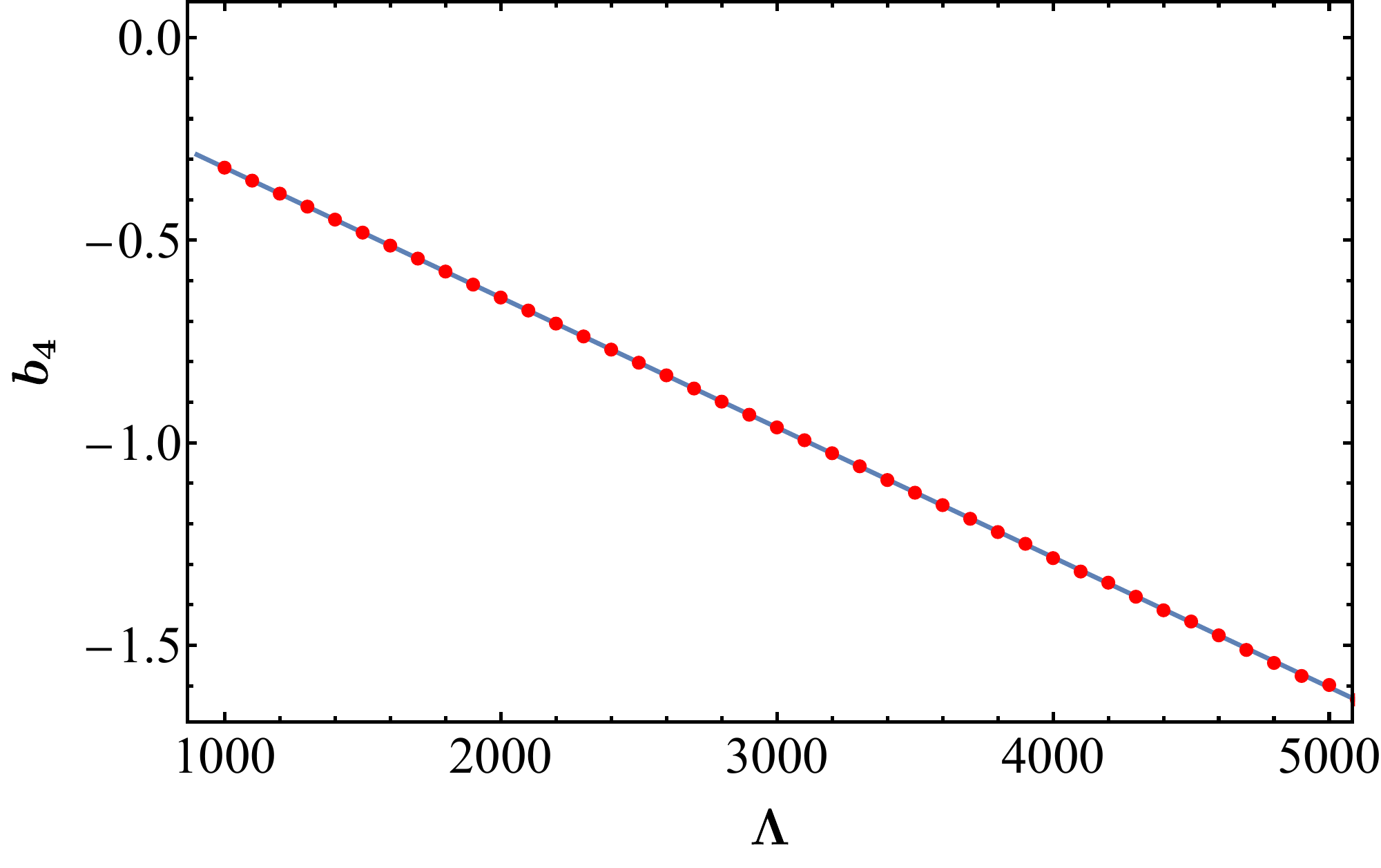}
	\caption{The red dots are the numerical values of the  coefficient of $q^4$, $b_4(\Lambda)$, for $R=1$ and $M=0.1$. The blue continuous line is the analytic prediction $b_4(\Lambda)=-{\Lambda}/{560 \pi ^{3/2}}$ contained in \eqref{V smooth expansions}.}
	\label{qpower4}
\end{figure}

The same kind of numerical analysis can also be performed exchanging the roles of $M$ and $q$. In the left panel of Fig.\,\ref{qpower2}, for instance, the red dots are the numerical values, for $R=1$ and $M=10$, of the coefficient of $q^2$ for values of $\Lambda$ within the range $10^3\le \Lambda\leq 6\times 10^3$, while the blue line is the analytic result $b_2(\Lambda)\equiv {\Lambda^3}/{80 \pi ^{3/2}}-{M^2 \Lambda}/{120 \pi ^{3/2}}$ contained in \eqref{V smooth expansions}. In the right panel we focus on a shorter range of $\Lambda$, namely $1.42\times 10^4\le \Lambda\le 1.45\times 10^4$, confronting the cases $M=0.1$ and $M=10^3$. The distance between the resulting two curves allows to  ascertain the presence of both terms ${\Lambda^3}/{80 \pi ^{3/2}}$ and $-{M^2 \Lambda}/{120 \pi ^{3/2}}$ in $b_2(\Lambda)$.
Another example is given in Fig.\,\ref{qpower4}, where the coefficient $b_4(\Lambda)$ of $q^4$ is initially plotted for $R=1$ and $M=0.1$. The {red} dots are the results of the numerical analysis, while the {blue} continuous line is the analytic prediction $b_4(\Lambda)=-{\Lambda}/{560 \pi ^{3/2}}$ contained in \eqref{V smooth expansions}. Successively we consider the case $M=1$, and obtain exactly the same plot, thus confirming the correspondence of the analytic and numerical results for $b_4(\Lambda)$, and its independence from $M$ (see above).
 

Finally, putting together all the results of the present section, we conclude that when a spherical smooth cutoff is used, $V_{1l} (\phi)$ is given by
\begin{equation}
	\label{V smooth}
	V_{1l}(\phi)=\frac{5 M^2+ 3q^2}{240 \pi ^{3/2}}R\Lambda ^3-\frac{35 M^4+14 q^2 M^2+3 q^4}{1680 \pi ^{3/2}}R\Lambda + \frac{M^5R}{60\pi} + \widetilde R_2 +\mathcal O\left(\Lambda^{-1}\right),
\end{equation}
which is nothing but \eqref{V smooth expansions} with the addition of the universal $M^5$ contribution. 

Equation \eqref{V smooth} is of paramount importance for our analysis. 
It shows that the UV-sensitive terms proportional to powers of $q$, whose presence in $V_{1l}(\phi)$ we first evidentiated using for the calculation a hard cutoff (see section 4.2), are also found when a smooth cut is used: the results \eqref{divpoti} and \eqref{V smooth} for $V_{1l}(\phi)$ (hard and smooth cutoff respectively) crucially differ from \eqref{divm} and \eqref{divm2} (the well-known UV-insensitive result of section 2) for the presence of additional UV-sensitive terms proportional to $q^2$ and $q^4$. Contrary to the largely diffused idea\cite{Contino:2001gz,Delgado:2001ex,Barbieri:2001dm,Masiero:2001im} that the appearance of these terms is due to the use of a hard cutoff, we have shown that when $V_{1l}(\phi)$ is calculated including the asymptotic regions of the virtual momenta in a consistent manner, UV-sensitive terms proportional to powers of $q$ show up, irrespectively of the use of a hard or smooth cutoff.    
Therefore, the comments that we have done below Eq.\,\eqref{divpotisusy} also apply to the present section.

Let us summarize the results obtained so far. In the usual approach (see sections 2, 4.1 and 5.1), where $V_{1l}(\phi)$ is calculated performing the infinite sum over $n$ independently from the integration over $p$, only divergences proportional to powers of $M$ (eventually cancelled in SUSY theories) appear. 
In sections 3, 4.2 and 5.2, however, we have shown that, when the calculation of $V_{1l}(\phi)$ is properly done, 
new field dependent UV-sensitive terms proportional to powers of $q$ show up. 
They originate from the presence of non-trivial boundary conditions on a multiply connected spacetime. These additional terms are artificially cancelled out when the asymptotic contribution of the fifth component of the loop momentum $p^{(5)}$ is included in $V_{1l}(\phi)$ independently from the asymptotic contributions of the other components.  
We have also found that the finite contribution to $V_{1l}(\phi)$ is always obtained when the physical conditions $q^2/\Lambda^2,M^2/\Lambda^2 \ll 1$ are fulfilled, and is physically relevant when the size of the compact dimension is sufficiently small (compared to the size of the $4D$ box).

Coming back to the smooth cutoff case of the present section, we point out that, using smooth cutoff functions different from ours, some authors recovered the UV-insensitive result for $V_{1l}(\phi)$. In the next section we investigate on this issue, considering three notable examples, namely the \vv thick brane"\cite{Delgado:2001ex}, the \vv Pauli-Villars"\cite{Contino:2001gz}, and the \vv proper time"\cite{Antoniadis:1997ic,Antoniadis:1997zg} calculations.

\section{Comparison with other smooth cutoff regularizations}
We now compare the results that we obtained in section 5.2, where for the calculation of $V_{1l}(\phi)$ we used the smooth cutoff function $e^{-(p^2+n^2/R^2)/\Lambda^2}$,    with those present in the literature, where other smooth functions are considered. We focus on three different realizations: \vv thick 
brane" \cite{Delgado:2001ex}, \vv Pauli-Villars" \cite{Contino:2001gz}, and \vv proper time" \cite{Antoniadis:1997ic,Antoniadis:1997zg}. To make contact with the existing literature, for the thick brane and  Pauli-Villars regularizations  we consider tadpole one-loop corrections $\delta m^2$ to the Higgs mass (rather than the one-loop corrections to the Higgs potential), that without regularization read
\begin{equation}
\label{tadpole unreg}
\delta m^2 =\sum_{n=-\infty}^\infty\int \frac{d^4p}{(2\pi)^4}\frac{1}{p^2+\left(\frac{n}{R}+q\right)^2+m^2}\,.
\end{equation}
When discussing the proper time, we will go back to $V_{1l}(\phi)$. 

For the thick brane case\cite{Delgado:2001ex}, the authors consider a model with orbifold compactification, and interactions localized on branes placed at the orbifold fixed points. Indicating with $z$ the coordinate along the compact dimension, these interactions in the $5D$ action $\mathcal S_{_{(5)}}$ are regularized by smoothening the localizing delta function $\delta(z)$ with a Gaussian distribution $f(z,l_s)=1/(\sqrt{2\pi}l_s)e^{-{z^2}/{l_s^2}}$, where $l_s$ is the thickness of the brane. 
Performing in $\mathcal S_{_{(5)}}$ the integration over $z$, the $4D$ action $\mathcal S_{_{(4)}}$ is generated, and the coupling constants turn out to be multiplied by $e^{-{(n/R+q)^2\pi^2}/{2\Lambda^2}}$, so that \eqref{tadpole unreg} becomes\footnote{In the model discussed in \cite{Delgado:2001ex} $m^2=0$. Eq.\,\eqref{thick brane} is more general, but this does not change the conclusions.} 
\begin{equation}
\label{thick brane}
\sum_{n=-\infty}^\infty\int^{\Lambda} \frac{d^4p}{(2\pi)^4}\frac{e^{-\frac{(n/R+q)^2\pi^2}{2\Lambda^2}}}{p^2+\left(\frac{n}{R}+q\right)^2+m^2}\,.
\end{equation}
The function $e^{-{(n/R+q)^2\pi^2}/{2\Lambda^2}}$ provides a smooth cut for the sum over the fifth component $p_5=n/R$ of the $5D$ momentum, while  the hard spherical cutoff $\Lambda \equiv l_s^{-1}$ is used for
the integration over the remaining four components (this is indicated by the uppercase $\Lambda$ in the above integral). Altogether,
the sum/integration over the $5D$ loop momentum $p^{(5)}$ is done with the cutoff $e^{-{(n/R+q)^2\pi^2}/{2\Lambda^2}} \theta(\Lambda^2-p^2)$.

From \eqref{thick brane} we see that the fifth component $p_5=n/R$ appears only in the combination $n/R+q$, both in the cutoff function and in the original propagator. 
From a physical point of view, such a combination in the cutoff function is disturbing, as for each of the $q_i$ that appear in the contribution to $\delta m^2$ it introduces a different cutoff function. On the contrary, our cutoff function $e^{-n^2/(R^2\Lambda^2)}$ does not depend on the $q_i$, so that in $\delta m^2$ the high energy modes are all uniformly cut. We will further comment on this point later.

Let us move to the Pauli-Villars regularization\cite{Contino:2001gz}. The authors consider the model introduced in \cite{Barbieri:2000vh}, and implement this regularization through the insertion in $\mathcal S_{_{(5)}}$ of higher derivative terms. The $5$-dimensional d'Alambertian $\Box_{(5)}$ is replaced by
\begin{equation}\label{box}
\Box_{(5)}(1+\Box_{(5)}^2/\Lambda^4)\,,
\end{equation}
so that Eq.\,\eqref{tadpole unreg} becomes\footnote{In the model considered in \cite{Contino:2001gz} the boson and fermion partners are considered together, and the loop integral for the two point function is then 
	\begin{align*}
		\sum_{n=-\infty}^\infty\int \frac{d^4x}{(2\pi)^4} x^2 \Bigg[\frac{(\Lambda R)^8}{\left[(\Lambda R)^4+\left(x^2+(2n)^2\right)^2\right]^2}\left(\frac{1}{x^2+(2n)^2}\right)^2
		-\frac{(\Lambda R)^8}{\left[(\Lambda R)^4+\left(x^2+(2n+1)^2\right)^2\right]^2}\left(\frac{1}{x^2+(2n+1)^2}\right)^2\Bigg],
	\end{align*}
where $x\equiv pR$ is the adimensional four-momentum, and $q$ takes the values $q=0,1/2$ for bosons and fermions respectively. The arguments developed in this section apply to each of the above integrals. }
\begin{equation}
	\label{pauli villars}
	\sum_{n=-\infty}^\infty\int \frac{d^4p}{(2\pi)^4}\frac{\Lambda^4}{\Lambda^4+\left(p^2+\left(\frac{n}{R}+q\right)^2\right)^2}\frac{1}{p^2+\left(\frac{n}{R}+q\right)^2+m^2}\,.
\end{equation}
From \eqref{pauli villars} we see that \eqref{box} generates the smooth cutoff function {\small ${\Lambda^4}\Big/{\Big[\Lambda^4+\left(p^2+\left(\frac{n}{R}+q\right)^2\right)^2\Big]}$} for the sum/integration over the $5D$ momentum $p^{(5)}$. Moreover, as for the thick brane case \eqref{thick brane}, $n/R$ appears only in the combination $n/R+q$, so that the observations we made above also apply to the present case. 

Pushing our analysis a step further, we proceed now to a comparison between the smooth cutoff function $e^{-{(n/R+q)^2\pi^2}/{2\Lambda^2}}$ for the sum over $n$ in \eqref{thick brane} and the corresponding function that we introduced in section 5.2, namely $e^{-{n^2/(R^2\Lambda^2)}}$ that appears in \eqref{pot smooth cutoff}. In section 5.2 we calculated the one-loop Higgs potential $V_{1l}(\phi)$, while here we are considering a tadpole contribution to the Higgs mass. But this is not a problem, since a typical tadpole contribution of the kind \eqref{tadpole unreg} emerges from our $V_{1l}(\phi)$ with each term of the sum multiplied by $e^{-{n^2/(R^2\Lambda^2)}}$. The former cutoff function can be obtained from the latter replacing\footnote{In \eqref{pauli villars} a different cutoff function appears, but it still depends on $q$ through the combination $n/R+q$, and the considerations that we develop below also apply to that case.} $n/R$ with $n/R+q$. 
Although such a replacement might seem a harmless deformation of our cutoff function, the final result strongly depends on whether one or the other of these functions is used. We will see in fact that (i) the presence of the combination $n/R+q$ in the cutoff function, and (ii) the fact that in\cite{Delgado:2001ex}
and\cite{Contino:2001gz} the sum over $n$ is  done {\it independently} from the integral over $p$, together conspire to make the final result artificially UV-insensitive: they realize a washing out of the UV-sensitive terms proportional to powers of $q^2$. 

To further investigate on this point, we begin by observing that both \eqref{thick brane} and \eqref{pauli villars} contain functions of the kind $f_{_\Lambda}(p,n/R+q)$, where the dependence on $n$ only comes through the combination $n/R+q$. Within the usual strategy (where the sum over $n$ and the integral over $p$ are performed independently from one another), we have then to calculate expressions of the kind
\begin{equation}
	\label{example}
	A(qR)\equiv \sum_{n=-\infty}^\infty C_{_\Lambda}\left(n +qR \right) 
\end{equation}
with
\begin{equation}\label{C}
C_{_\Lambda}(n+qR) \equiv	\int \frac{d^4p}{(2\pi)^4} \,\,f_{_\Lambda}\left(p,\frac{n}{R}+q\right)\,.
\end{equation}
For the purposes of the present discussion, we are only interested in the presence or absence of divergences proportional to powers of $q$, so we  disregard the possible presence of other divergent terms (that in SUSY theories in any case disappear when the contributions of each couple of boson and fermion superpartners are combined).

The right hand side of \eqref{example} can be evaluated resorting again to the EML formula 
\begin{align}\label{emlmassa}
	A(qR)&=\int_{-\infty}^\infty dx\, C_{_\Lambda}(x+qR) \nn \\ 
	&+\lim_{L\to\infty}\Bigg(\frac {C_{_\Lambda}(L+qR)+C_{_\Lambda}(-L+qR)}{2}    +B_2\frac{C^{(1)}_{_\Lambda}(L+qR)-C^{(1)}_{_\Lambda}(-L+qR)}{2}+R_{2}\Bigg),
\end{align} 
where $C^{(1)}_{_\Lambda}(x)$
indicates the first derivative of the function $C_{_\Lambda}(x)$, $B_2$ is the Bernoulli number $B_i$ with $i=2$, and $R_2$ the rest. The integer $L$ in \eqref{emlmassa} is needed in the intermediate steps of the calculation, and is eventually sent to infinity. Inserting \eqref{emlmassa} in  \eqref{thick brane} and \eqref{pauli villars}, we find that in both cases all the terms in the second line of \eqref{emlmassa}, with the exception of $R_2$, are $\mathcal O(L^{-1})$, so they vanish in the $L \to \infty$ limit. In the same limit $R_2$ gives the usual result, with its well-known finite periodic dependence on $q$. Therefore, the only possible source of ($q$-dependent) divergences in \eqref{emlmassa} is the integral. However, due to the choice of a cutoff function that depends on $n$ only through the combination $n/R+q$, the trivial change of variable $x\to x+qR$ shows that no $q$-dependent terms arise from it, and a fortiori no $q$-dependent divergences. 

To better appreciate the difference with our smooth cutoff function $e^{-n^2/(R^2\Lambda^2)}$, we go back to our result \eqref{pot smooth cutoff} for $V_{1l}(\phi)$ and consider the derivation of a tadpole contribution to the mass. A simple inspection of \eqref{pot smooth cutoff} shows that in this case Eq.\,\eqref{example} is replaced by an expression of the kind
\begin{equation}
	\label{examplemod}
	\widetilde	A(qR)\equiv \sum_{n=-\infty}^\infty \widetilde C_{_\Lambda}\left(n +qR;n \right) \,,
\end{equation}
where the terms of the series do not depend only on the combination $n+qR$, but also on $n$ alone.

This difference between  $ A(qR)$  and  $\widetilde A(qR)$ is crucial. We have already argued that from the physical point of view $A(qR)$ presents the drawback that it generates different cutoff functions for different values of $q$ in the same potential, while $\widetilde A(qR)$ results from a cutoff function that implements the cut in the high energy modes in a physically uniform way for all the terms that contribute to the potential. Now, applying the EML formula to \eqref{examplemod}, we see that in the integral  $\int_{-\infty}^\infty dx\, \widetilde C_{_\Lambda}(x+qR;x)$ it is not possible to get rid of the dependence on $q$ by performing the change of variable $x+qR \to x$. These observations allow us to understand the reason why in the one-loop Higgs potential $V_{1l}(\phi)$ in \eqref{V smooth}, that is calculated with the smooth cutoff function $e^{-n^2/(R^2\Lambda^2)}$, UV-sensitive terms proportional to powers of $q^2$ appear. 

Differently from the UV-sensitive terms proportional only to powers of $M^2$, the $q^2$-dependent ones do not disappear even when supersymmetric theories are considered. 
They are profoundly different from the former, and are the hallmark of the UV-sensitivity of $V_{1l}(\phi)$ that comes from the non-trivial boundary conditions allowed by the compact extra dimensions when the spacetime manifold is multiply-connected (see also the discussion below Eq.\,\eqref{divpotisusy}). 

We can ask why, when considering the thick brane and Pauli-Villars regularizations, smoothening functions that depend on the combination $n/R+q=p_5+q$ appear. The reason is that these functions come out from deformations of the original lagrangian in $x$-space (see comments above Eqs.\,\eqref{thick brane} and \eqref{pauli villars} for thick brane and Pauli-Villars respectively), so that the non-monodromies of the fields along the compact dimension (see \eqref{hosotani}) generate the quantity $p_5+q$ for the regulating functions in momentum space. This results in cuts on $n/R+q$ rather than on the fifth component of the momentum $p_5=n/R$.
But we already stressed that this is physically not sound, as different cutoff functions are generated for each value of the charges $q_i$. 
Therefore, the thick brane and Pauli-Villars regularizations implement an \vv ad hoc'' procedure that artificially cancels the divergences proportional to powers of $q^2$. In other words, a class of specially chosen functions forces the calculations to wash out UV-sensitive terms that are actually present. 
There is another reason why such a way of implementing cutoff functions is physically unacceptable. For models with an effective potential of the kind \eqref{EP2}, where $q=q(\phi)$, a field dependence in the cutoff function would appear, but clearly this is physically unacceptable.

We now move to consider the calculation of $V_{1l}(\phi)$ with the proper time regularization\cite{Antoniadis:1997ic,Antoniadis:1997zg}. Writing the loop integral with the help of the Schwinger identity,
replacing 
 the lower extreme \vv $0$" for the integration over $s$ with $1/\Lambda^2$, and performing the $s$ integral, we get 
\begin{align}
	\label{proper time 1}
	V_{1l}(\phi)&=-\sum_{n=-\infty}^\infty\int \frac{d^4p}{(2\pi)^4} \int_{\frac{1}{\Lambda^2}}^\infty\frac{ds}{s} \left\{e^{-s\left(p^2+M^2+\left(\frac{n}{R}+q\right)^2\right)}-e^{\left(p^2+\frac{n^2}{R^2}\right)}\right\} \nonumber \\
	&= -\sum_{n=-\infty}^\infty\int \frac{d^4p}{(2\pi)^4} \left\{\Gamma \left(0,\frac{p^2+M^2+\left(\frac{n}{R}+q\right)^2}{\Lambda ^2}\right)-\Gamma \left(0,\frac{\frac{n^2}{R^2}+p^2}{\Lambda ^2}\right)\right\} 
\end{align}
where $\Gamma(x,y)$ is the incomplete gamma function. 
The proper time regularization then casts the calculation of $V_{1l}(\phi)$ in the form 
\begin{equation}
	\label{example 2}
	V_{1l}(\phi)= \sum_{n=-\infty}^{\infty}	\int \frac{d^4p}{(2\pi)^4} \left(f_{_\Lambda}\left(p,\frac{n}{R}+q\right)+g_{_\Lambda}\left(p,\frac nR\right)\right),  
\end{equation}
that is of the kind \eqref{example} (disregarding the second term in the right hand side of \eqref{example 2} that is related to the subtraction of field independent quantities, and then irrelevant for our physical considerations). The comments made above for the thick brane and Pauli-Villars regularizations then also apply to this case. 

Before ending this section, we stress again that
the above attempts to implement smooth cutoff regularizations were made to overcome the objections originally raised in\cite{Ghilencea:2001ug} against the UV-insensitivity (finiteness) of $V_{1l}(\phi)$.
Our conclusion is that they are flawed by the use of cutoff functions that are physically unacceptable and artificially realize the cancellation of UV-sensitive terms. On the contrary, the use of physically sound cutoff functions shows that these are genuine UV-sensitive contributions that have to be taken into account. 

\section{Summary and conclusions}
We considered higher-dimensional theories with compact extra dimensions and studied the UV-sensitivity of the Higgs effective potential $V_{1l}(\phi)$ derived from these models.
About twenty years ago, some  results indicated that $V_{1l}(\phi)$, as well as the Higgs boson mass $m_H$, could be UV-insensitive (Introduction and section 2). The possibility of having a finite $m_H$ 
was clearly seen as a very welcome result, leading to the widespread belief that compactification could be combined with unbroken higher-dimensional supersymmetry to alleviate (and possibly get rid of) the naturalness problem, while still furnishing at low energies the typical spectrum of a softly broken SUSY theory. 
Objections against  these results were raised, that triggered a heated debate. However, the community soon came (or seemed to came) to a general agreement in favour of the correctness of the finite results for $V_{1l}(\phi)$, that have since been used in many different applications, even in very recent times.

The search for a mechanism, a symmetry, a theoretical framework, where the Higgs mass could show UV-insensitivity, or at least a much milder sensitivity to UV physics than the typical one, is clearly a question of the greatest importance. 
This motivated us to reconsider the entire approach based on higher-dimensional theories with compact extra dimensions.
Combining analytical and numerical methods, we found that the UV-insensitivity of the Higgs one-loop potential $V_{1l}(\phi)$ is illusory. In fact, it results from physically and mathematically illegitimate steps in the calculation, and  we have shown in detail how an artificial cancellation of UV-sensitive terms occurs.
For the purposes of our analysis, it was sufficient to consider $5D$ theories with one compact dimension in the shape of a circle of radius  $R$ (sections 3, 4 and 5), but the results can immediately be extended to more general cases. They show that $V_{1l}(\phi)$ has a previously overlooked UV sensitivity, whose physical origin is in the boundary conditions of the fields in spacetimes with non-trivial topology.

We have found that the usual result of a UV-insensitive Higgs potential $V_{1l}(\phi)$ comes from an incorrect treatment of the loop momentum  asymptotics in the calculation: the fifth component $p_5$ of the loop momentum $p^{(5)}$ is included separately from the asymptotic contributions of the other components.  
In sections 4.1 we have explicitly shown how such an artificial cancellation of the above mentioned UV-sensitive terms from $V_{1l}(\phi)$ takes place in the original calculation. In sections 3, 4.2 (with a hard cutoff) and 5.2 (with a smooth cutoff) we have shown the way to properly include the asymptotics of the loop momentum in the calculation of $V_{1l}(\phi)$. As a result, we found the UV-sensitive terms previously overlooked.  

In section 6 we analysed certain smooth regularizations considered in the previous literature that allowed to recover the finite result, and thus consolidated the belief that in $V_{1l}(\phi)$ and $m_H$ no UV-sensitive terms are present. We have shown that these calculations once again implement an artificial washing out of the UV-sensitive terms. 
In particular, this is what happens in the framework of the proper time regularization. This latter observation is relevant in connection with the usual way string theory calculations are related to field theory ones. In fact, unfolding the fundamental domain of the torus on which the string partition function is calculated, an integral that strongly resembles a proper time realization of a loop integral is obtained. Our results raise some warnings on the use the proper time integral as the appropriate bridge towards field theories with non-trivial boundary conditions\cite{Abel:2021tyt,Ghilencea:2001bv}.  

We cannot conclude the present work without discussing some well-known attempts to justify the asymmetrical treatment of the sum over $n$ and the integration over $p$. One is rooted in the comparison of this usual way of performing the calculation of $V_{1l}(\phi)$ with the corresponding one in finite temperature field theory for the free energy. It is pointed out that the infinite sum over $n$ has to be performed before the integration over $p$ as it is the case in finite temperature field theory for the sum over the Matsubara frequencies $\omega_n$. However, such an advocated analogy between these two cases is misleading. In fact, the sum over $n$ in finite temperature field theory is needed to realize the ensemble average, that implements the ergodic hypothesis, and as such must be extended up to infinity (otherwise ergodicity would be violated). The sum over $n$ and integration over $p$ in our case realize the inclusion in the theory of the quantum fluctuations, and obviously none of the components of the loop momentum can be treated differently from the others. In particular, we should be extremely careful when treating the asymptotics of each of these components\footnote{In this respect we note that we should not be misled by the fact that in finite temperature field theory in the imaginary time formalism the \vv space" is $\mathbb R^3\times S^1$, and in our case it is $\mathbb R^4\times S^1$. We stress again that the crucial difference between the two cases, is that in the latter case we cannot send $n\to \infty$ otherwise we would treat inconsistently the different components of the loop momentum, while in the former case we must send $n\to\infty$  otherwise we would violate ergodicity.}.      

Another apparently different attempt consists in performing the calculation of the two-point Green's function in the \vv mixed position-momentum" space\cite{Arkani-Hamed:2001jyj}. This corresponds to a physically unrealizable separation of the dynamics along the compact dimension ($S^1$ in our case) from that in $\mathbb R^4$. Actually, trying to solve the dynamics along the circle while keeping the other components frozen (i.e. technically swapping $\partial^2_{(4)}\leftrightarrow p^2_{(4)}$ and treating $p^2_{(4)}$ as a constant in the equations of motion), is tantamount to perform the infinite sum over $n$ first, and is equally unacceptable on physical grounds.    

We conclude underlining once again that the deep physical reason for the appearence of UV-sensitive terms in $V_{1l}(\phi)$ is in the non-trivial boundary conditions allowed by the multiply connected nature of the spacetime. We do not see how such a physical effect related to the non-trivial topology of the spacetime could be circumvented to make these terms disappear from $V_{1l}(\phi)$.

\section*{Acknowledgements}
We would like to thank Giuseppe Di Fazio, Ivano Lodato and Giovanni Russo for helpful discussions. The work of CB is supported by Basic Science Research Program through the National
Research Foundation of Korea (NRF) funded by the Ministry of Education, Science and Technology (NRF-2022R1A2C2003567). The work of VB and FC is carried out within the INFN project QFT-HEP.

\section*{Appendix}
In Eqs.\,\eqref{thick brane}, \eqref{pauli villars} and \eqref{proper time 1} in the text we have to perform the sum and the integral of a function of the kind $f_{_{\Lambda}}(p,n/R+q)$, where the dependence on $n$ comes only through the combination $n/R+q$ (apart from a possible field-independent normalization). Actually $f_{_\Lambda}$ depends on the square of $n/R+q$, so that we have to calculate an expression of the kind
\begin{equation}
	\label{examplecopia}
	F(qR)\equiv \sum_{n=-\infty}^{\infty}\left(	\int \frac{d^4p}{(2\pi)^4} \left[f_{_\Lambda}\left(p,(n+qR)^2; \Lambda\right) + {g_{_\Lambda}\left(p,n/R; \Lambda\right)}\right] \right). 
\end{equation}

In the text we have found that the function $F(qR)$ written above is periodic (see \eqref{Vrx}, polylogarithmic functions). Interestingly, we can show that the fact that a function $F(qR)$ of the kind defined in \eqref{examplecopia} is periodic, with period the unitary interval, can be seen even before perfoming the actual calculation.
In fact, being the sum infinite, when $q R\in \mathbb N$ the final result for $F(qR)$ does not depend on $q$ (we can operate the replacement $n+qR \to n$ in the summation index). This means that a $q$-dependence can only arise from values of $q$ such that $q R\,\,\cancel{\in} \,\,\mathbb N$. 
From \eqref{examplecopia} we immediately see that $F(qR)=F(-qR)$ and that $F(qR+s)=F(qR)$, where $s$ is an integer. Combining the two, we also have that $F(qR)=F([qR]+1-qR)$, where $[qR]$ is the integer part of $qR$. There are two important consequences of these relations:  (i) the function $F(qR)$ is periodic, with period equal to the unitary interval; (ii) in each unitary interval $[0,1]$, $[1,2]$, \dots, the function $F(qR)$ is such that in the second half of the interval it is the mirror symmetric of itself in the first half of the interval.

We then see that, even before performing the actual calculation, $F(qR)$ can have only a periodic and oscillatory dependence on $q$ when the sum over $n$ is extended up to infinity.
To obtain such a result two crucial steps are needed: (i) the introduction of a regularizing function that casts the ill-defined original one-loop correction into a well-defined (convergent) contribution of the form \eqref{examplecopia} (\eqref{example 2} in the text); (ii) the infinite sum over $n$ is performed independently of the integration over the four-momentum. In this respect, we observe that even the one-loop potential considered in section 2 has the general form \eqref{examplecopia}, and that also for that calculation we followed the usual strategy of performing the infinite sum over $n$ independently from the integration over $p$. Therefore, all the considerations on the dependence on $q$ developed above apply straightforwardly to that calculation, so that the oscillatory form of $V_{1l}(\phi)$ had to be expected even in \eqref{litium}. 

Finally we note that in all the cases considered: (i) $\lim_{\Lambda\to\infty} f(p,n/R+q;\Lambda)\equiv h(p,n/R+q)$ is finite; (ii) $\sum_{n=-\infty}^\infty h(p,n/R+q)$ is convergent. As the argument developed above do not depend on the specific value of $\Lambda$, we conclude that the periodic function of $q$ does not contain $\Lambda$.

\end{document}